# Coherent Exciton-Lattice Dynamics in a 2D Metal Organochalcogenolate Semiconductor

Eric R. Powers[1,4], Watcharaphol Paritmongkol[1,2,4], Dillon C. Yost[3], Woo Seok Lee[1,3], Jeffrey C. Grossman[3], William A. Tisdale[1]*

[1]Department of Chemical Engineering, Massachusetts Institute of Technology, Cambridge, Massachusetts 02139, United States

[2]Department of Chemistry, Massachusetts Institute of Technology, Cambridge, Massachusetts 02139, United States

[3]Department of Materials Science and Engineering, Massachusetts Institute of Technology, Cambridge, Massachusetts 02139, United States

[4]These authors contributed equally.

*Correspondence: tisdale@mit.edu

**Current affiliations**:

W.P. – Department of Materials Science and Engineering, School of Molecular Science and Engineering, Vidyasirimedhi Institute of Science and Technology (VISTEC), Rayong 21210, Thailand

D.C.Y. – Design Physics Division, Lawrence Livermore National Laboratory

## Summary

Hybrid organic-inorganic nanomaterials can exhibit transitional behavior that deviates from models developed for all-organic or all-inorganic materials systems. Here, we reveal the complexity of exciton-phonon interactions in a recently discovered 2D layered hybrid organic-inorganic semiconductor, silver phenylselenolate (AgSePh). Using femtosecond resonant impulsive vibrational spectroscopy and non-resonant Raman scattering, we measure multiple hybrid organic-inorganic vibrational modes and identify a subset of these modes that strongly couple to the electronic excited state. Calculations by density functional perturbation theory show that these strongly coupled modes exhibit large out-of-plane silver atomic motions and silver-selenium spacing displacements. Moreover, analysis of photoluminescence fine-structure splitting and temperature-dependent peak-shifting/linewidth-broadening suggests that light emission in AgSePh is most strongly affected by a compound 100 $cm^{-1}$ mode involving the wagging motion of phenylselenolate ligands and accompanying metal-chalcogen stretching. Finally, red-shifting of vibrational modes with increasing temperature reveals a high degree of anharmonicity arising from non-covalent interactions between phenyl rings. These findings reveal the unique effects of hybrid vibrational modes in organic-inorganic semiconductors and motivate future work aimed at specifically engineering such interactions through chemical and structural modification.

## Introduction

Hybrid organic-inorganic materials can exhibit functionality and performance not achievable in all-organic or all-inorganic material systems alone. These hybrid materials – including metal-organic frameworks, colloidal nanocrystals, organics in sol-gel derived matrices, organic modified ceramics, and hybrid halide perovskites – have been used in many fields ranging from catalysis, photonics, functional coatings, electronics, energy, and sensing to biology, medicine, and biotechnology[1–3].

Interest in hybrid organic-inorganic semiconductors has advanced recently due to remarkable optoelectronic properties and ease of fabrication by low-temperature solution-based syntheses[3–6]. However, a complete description of excited state dynamics in hybrid semiconductors has proven elusive due to strong charge carrier-lattice interactions[7–11] that defy conventional models of all-inorganic semiconductors. Moreover, excited state dynamics in low-dimensional hybrid semiconductors are further complicated by the mixed Frenkel-Wannier nature of strongly-bound excitonic states[12–14].

Exciton-phonon coupling in two-dimensional (2D) hybrid semiconductors has been extensively studied in 2D hybrid organic-inorganic lead halide perovskites[12,15–17]. Previous research in this material class has shown that the main vibrational modes that couple strongly to excitons and excited electronic states belong to the motions of inorganic frameworks, while the organic components indirectly affect the vibrational modes through structural templating[15–17]. However, it is unclear whether these findings are representative of all 2D hybrid semiconductors or if they are specific only to 2D lead halide perovskites, which are uniquely ionic in nature and possess relatively weak hydrogen-bonding interactions between organic and inorganic components.

Layered metal organochalcogenolates (MOCs)[18,19] are an emerging class of covalently-bonded 2D hybrid organic-inorganic semiconductors with potential applications in light emission[20–22], thermoelectricity[23], and electrocatalysis,[24] as well as in light[25–27], chemical[28], and thermal[29] sensing. A prototypical member of this material family is 2D silver phenylselenolate (AgSePh), comprised of a hybrid quantum-well structure[30–32] which exhibits a thickness-independent direct bandgap[33], strong exciton binding energy[34,35], fast picosecond photoluminescence lifetime[26,31,34,36,37], narrow-linewidth blue luminescence centered at 467 nm[38], and 2D in-plane exciton anisotropy with polarized light absorption and emission.[35] Moreover, synthesis of AgSePh can be accomplished without specialized equipment, and the material can be prepared as single crystals[30,31], microcrystals[31,39,40], or polycrystalline thin films[26,33,41] *via* low-temperature vapor-phase or solution-phase processes, depending on the end-use need. These attributes, combined with earth-abundant elemental composition and chemical robustness, have generated interest in AgSePh for a variety of optoelectronic applications[25,27]. Recent experimental results also indicate the presence of strong exciton-phonon coupling in AgSePh[42].

In this work, we employ a combination of steady-state and time-domain spectroscopy techniques together with density functional perturbation theory (DFPT) to develop a detailed description of exciton-lattice interactions in AgSePh. Understanding of exciton-phonon coupling in low-dimensional semiconductors like AgSePh reveals new opportunities for controlling optoelectronic properties through molecular engineering of hybrid organic-inorganic systems.

## Results and Discussion

**Crystal Structure, Electronic Structure, Synthesis, and Basic Optical Properties of AgSePh**

AgSePh is a 2D hybrid organic-inorganic semiconductor, consisting of Ag, Se, C, and H. It crystallizes in the *C*2/*c* or *P*2$_1$/*c* space group[30–32] and adopts a quantum-well structure (Figure 1a) with a layer of Ag atoms located between two layers of Se atoms that are covalently bonded to phenyl (Ph) rings. Each Ag atom is surrounded by four Se atoms in a tetrahedral configuration, and each Se atom is linked to four Ag atoms and one phenyl ring oriented in the out-of-plane direction. The Ag atoms are arranged into a distorted hexagonal pattern (Figure 1b), leading to in-plane anisotropy[35] and three different Ag-Ag separation distances (2.99, 2.90 and 3.03 Å) for the *P2$_1$/c* crystal structure. We refer to these three Ag-Ag separations as (Ag-Ag)$_1$, (Ag-Ag)$_2$, and (Ag-Ag)$_3$, respectively, as defined in Figure 1b. The short Ag-Ag bond distances (shorter than the sum of two Ag atomic radii, 3.44 Å) suggests strong argentophilic interactions within the 2D plane.[32,43]

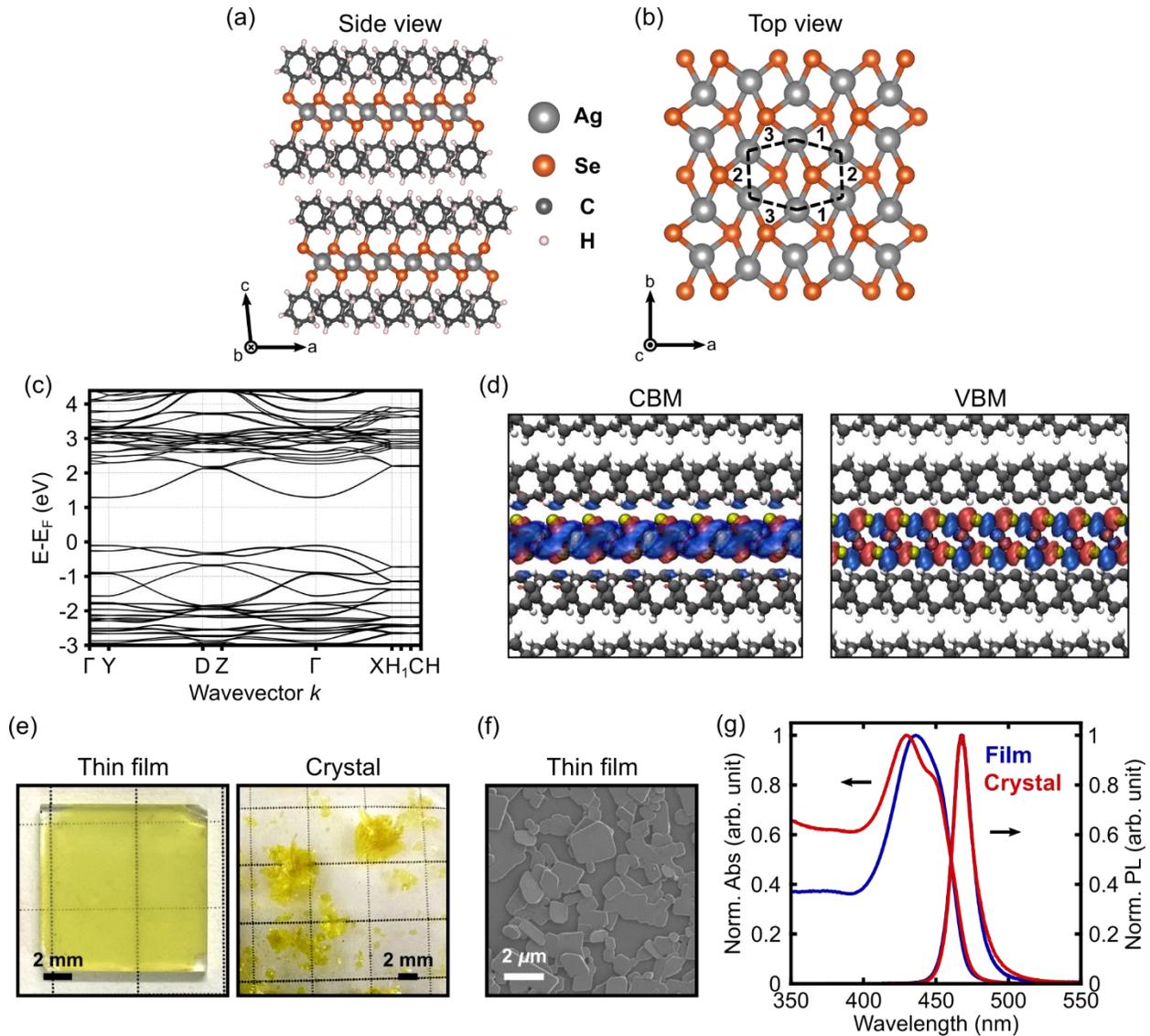

**Figure 1. Structural, electronic, and optical properties of AgSePh.** (a) Structure of AgSePh when viewed from the side, showing a natural 2D quantum-well system. (b) Structure of AgSePh *P2₁/c* space group when viewed from the top, showing the distorted hexagonal pattern of in-plane Ag atoms and three different Ag-Ag bond lengths labeled by 1, 2, and 3. (c) Calculated electronic band structure of AgSePh derived from density functional theory, predicting a direct bandgap at Γ. An image of the corresponding Brillouin zone is given in Figure S1. (d) Isosurfaces of the wavefunctions at the conduction band minimum (CBM) and valence band maximum (VBM) at Γ. (e) Photos of an AgSePh thin film and AgSePh crystals used in this study. (f) Scanning electron micrograph of an AgSePh thin film showing its polycrystalline morphology with random in-plane orientation. (g) Absorption and photoluminescence spectra of AgSePh thin films and crystals (powder).

Electronic structure calculations performed using density functional theory (DFT) show that AgSePh is a semiconductor with a direct bandgap at Γ and no dispersion along the out-of-plane direction, in agreement with its quantum-well structure and a previous calculation[35] (Figure 1c). Figure 1d shows the electron densities at the conduction band minimum (CBM) and valence band maximum (VBM), which are mainly concentrated in the inorganic AgSe core. Electron density at the CBM also contains a small orbital contribution from the nearest C atoms, suggesting the possibility for electronic bandgap tunability by organic modification.

AgSePh was prepared as both thin films and large crystals, based on previously reported procedures.[26,31] AgSePh thin films (Figure 1e, left) were prepared by vapor-phase chemical transformation – or "tarnishing" – of metallic silver films by diphenyl diselenide ($Ph_2Se_2$) in the presence of dimethyl sulfoxide (DMSO) vapor.[26] A scanning electron micrograph of a representative film shows its polycrystalline nature with micrometer grain sizes and random in-plane orientation on a glass substrate (Figure 1f). Using an alternate synthetic method, large AgSePh crystals (Figure 1e, right) were obtained by an organic single-phase reaction between silver nitrate ($AgNO_3$) and $Ph_2Se_2$ in a mixed propylamine and toluene solution.[31]

Figure 1g shows the absorption and photoluminescence spectra of AgSePh thin films and crystals, revealing a crowded excitonic absorption feature at 430-450 nm comprised of three distinct optical transitions, and a single photoluminescence peak at 467 nm, in agreement with previous reports[26,31,35,37,38]. AgSePh thin films were used for femtosecond pump-probe experiments, while crystals were used for non-resonant Raman and photoluminescence spectroscopy experiments.

**Time-domain observation of coherent exciton-lattice dynamics by impulsive vibrational spectroscopy (IVS)**

To investigate vibrational dynamics in AgSePh, we employed resonant impulsive stimulated Raman scattering (RISRS)[9,10,12,15,16], a time-domain vibrational spectroscopy technique belonging to a broader class of methods known collectively as impulsive vibrational spectroscopy (IVS). [44–48] Based on pump-probe transient absorption spectroscopy, IVS measures the change in a sample's absorption/transmission due to photoexcitation as a function of time (Figure 2a). In our experiment, an ultrafast 375 nm pump laser pulse was used to excite AgSePh electronically and vibrationally, followed by a broadband probe laser pulse to monitor the excited state dynamics over a controllable time delay (Figure 2b). Due to the short temporal width (~70 fs) of the pump laser pulse (Figures S2 and S3), a coherent vibrational wavepacket is generated on an excited-state potential energy surface upon photoexcitation, and this wavepacket evolves over time with a frequency ω corresponding to the underlying vibrational mode(s). This oscillation of the wavepacket leads to fluctuations in the atomic spacing and lattice organization, resulting in a time-dependent modulation of sample's absorption that contains information on the sample's vibrational dynamics.

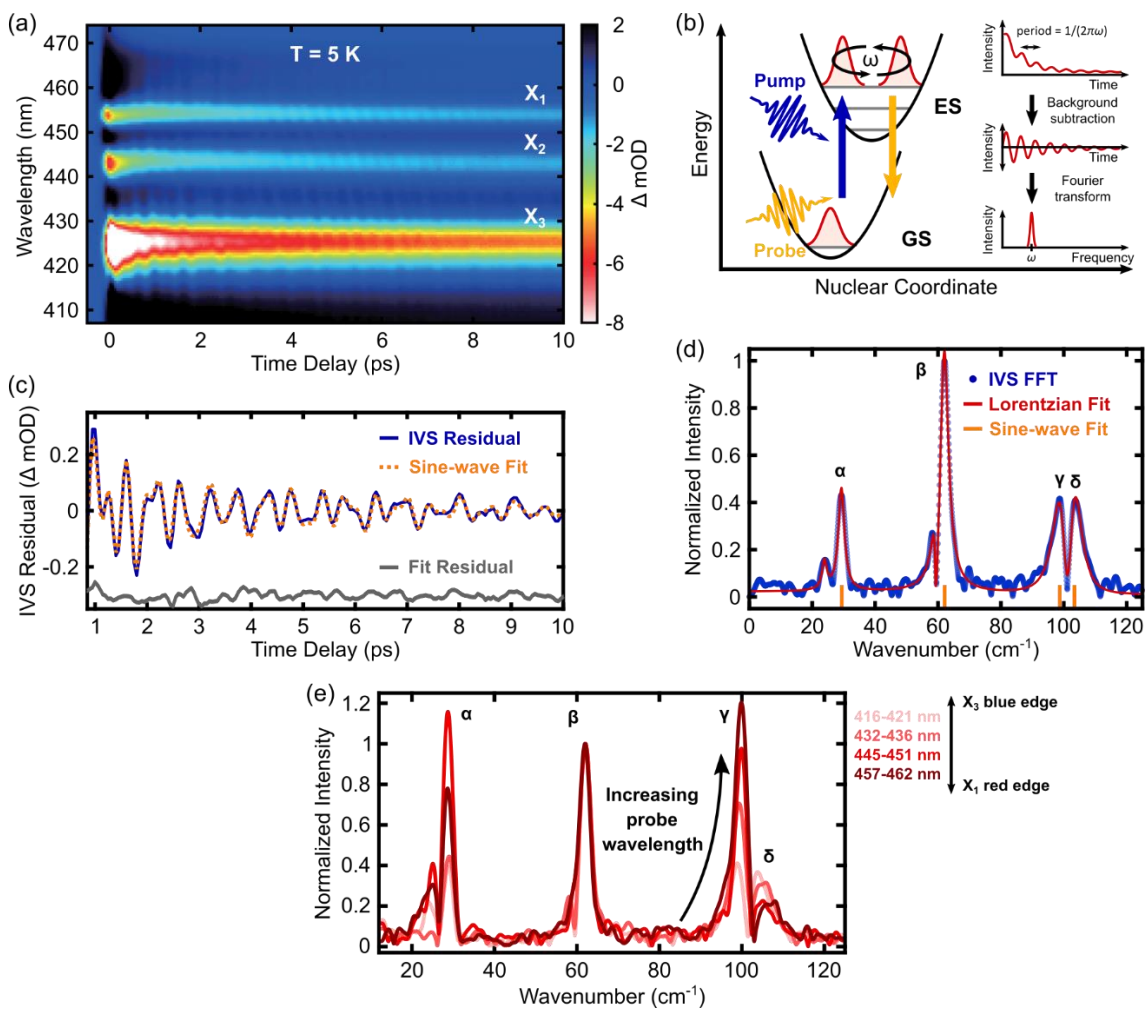

**Figure 2. Coherent vibrational dynamics obtained by impulsive vibrational spectroscopy (IVS).** (a) IVS color map of thin-film AgSePh at 5 K. The IVS data were collected and analyzed over a time delay of 0 to 20 picoseconds (ps), at which point all substantial oscillatory signals were found to have degraded. The data are truncated to 10 ps here for clarity; the full 20 ps color map can be found in Figure S11a. The three distinct excitonic transitions $X_1$, $X_2$, and $X_3$ are labeled.

(b) Schematic of the IVS technique employed in a displaced harmonic oscillator potential energy system. Photoexcitation occurs from the electronic ground state (GS) to the excited state (ES) and induces a coherent vibrational wavepacket oscillating at a frequency ω. At right, the process of extracting vibrational frequency information from the IVS signal is illustrated, involving the subtraction of fitted electronic dynamics to obtain vibrational dynamics, followed by a Fourier transform to find the vibrational mode frequency. (c) Analysis of time-domain IVS data, showing the vibrational oscillations in the signal over a 408-422 nm region (blue), the fit to multiple decaying sine waves (orange), and the difference between the experimental signal and fit with an offset for clarity (gray). (d) Vibrational frequencies obtained after Fourier transformation of IVS vibrational dynamics (blue). These frequency-domain data were fit to multiple Lorentzian peaks (red) and agree with the frequencies identified by the sine-wave fit to the time-domain data (orange ticks). (e) Probe wavelength-dependent IVS vibrational frequencies showing an increase in the relative intensity of the γ mode at longer probe wavelengths.

Figure 2a shows the IVS color map for AgSePh at 5 K, plotted as the change in absorption versus time delay and probe wavelength. We observed three prominent bleach features (a negative change in absorption), labeled $X_1$, $X_2$, and $X_3$.[35,36,42] These three optical transitions in AgSePh were previously assigned to three distinct excitonic states oriented within the 2D plane ($X_2$ perpendicular to $X_1 + X_3$) and exciton binding energy ≥300 meV[35]. A fit to these excitonic dynamics and further discussion are shown in Figure S4.

Superimposed on the decaying electronic signals in the IVS data are the oscillatory vibrational dynamics. To isolate the vibrational dynamics, we sum over a wavelength region of the IVS data, then fit and subtract off the electronic contribution to the signal (Figures 2b and S5), leaving only the quickly oscillating vibrational signal as shown in Figure 2c. Next, a Fourier transform is applied to extract the frequencies of underlying vibrational modes (see Supplementary Note 2 for an extended discussion on this data analysis method). Figure 2d shows the vibrational frequencies of the IVS signal at 408-422 nm.

We observe four dominant vibrational modes in the IVS data (Figure 2d) with frequencies of 29.3, 62.1, 98.7, and 103.7 cm$^{-1}$ labelled α, β, γ, and δ, respectively. These are the vibrational modes that significantly modulate the excitonic optical transitions in AgSePh; or, in other words, these are the primary modes that participate in exciton-phonon coupling.

The IVS results show differences in exciton-phonon coupling behavior among the three excitonic states (Figure 2e). While the β mode is the most dominant mode based on signal intensity at probe wavelengths of 416-421 nm (corresponding to the highest-energy excitonic state, $X_3$), the $\gamma$ mode is most intense for the lowest-energy excitonic state, $X_1$ (probe wavelength 457-462 nm).

To corroborate the findings of the frequency-domain analysis, we also performed mode analysis directly on the time-domain data (Figure 2c) by fitting to the sum of multiple decaying sine waves of the form

$$I = \kappa_1 \sin(\omega_1 t + \varphi_1) \exp\left(\frac{-t}{\tau_1}\right) + \kappa_2 \sin(\omega_2 t + \varphi_2) \exp\left(\frac{-t}{\tau_2}\right) + \cdots, \qquad (1)$$

where $I$ is the intensity of the IVS signal, $t$ is time, and the remaining variables represent adjustable fitting parameters. The result of a four-mode fit is shown as an orange dashed line overlaid with the original data in Figure 2c; details on the fitting method and a list of the derived fitting parameters can be found in Supplementary Note 3 and Table S4, respectively. The orange ticks along the x-axis in Figure 2d represent the fitted sine-wave frequencies, showing good agreement with the frequencies obtained by Fourier transformation. Beyond identification of the vibrational frequencies, the unique time-domain data collection method of IVS allows for analysis of the temporal evolution of the vibrational coherences, represented by the $\tau_i$ coherence lifetime terms in the fit and tabulated in Table 1.

**Comparison of IVS spectrum to non-resonant Raman scattering (NRRS)**

The unique information content of IVS is further revealed by comparison of the IVS spectrum to the corresponding frequency-domain non-resonant Raman scattering (NRRS) spectrum (Figure 3a). By using a non-resonant Raman pump laser wavelength of 785 nm (below the bandgap of AgSePh), we ensure that only vibrational modes belonging to the electronic ground state are observed in NRRS.

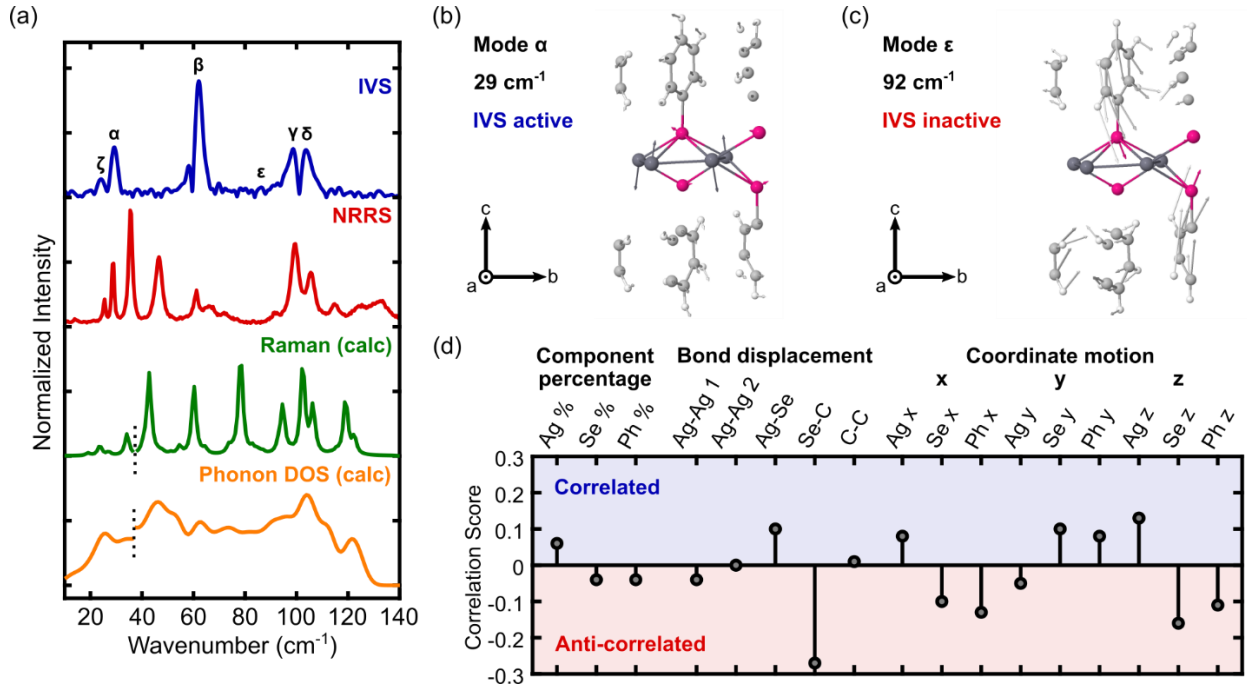

**Figure 3. Vibrational mode identification using density functional perturbation theory (DFPT).** (a) Comparison of vibrational spectra from several experiments and simulations, normalized and offset. Shown in order are impulsive vibrational spectroscopy (IVS, blue, 5 K), non-resonant Raman scattering (NRRS, red, 78 K), DFPT calculated Raman (green), and DFPT calculated phonon density of states (orange). See also Figures S10, S15, and S16. For the spectral range above 37 cm$^{-1}$ (marked by a black dotted line), the calculated Raman spectrum and phonon density of states were shifted by 8 cm$^{-1}$ to better match the experimental spectra. (b and c) DFPT simulated atomic displacements in the AgSePh structure for the α (b) and ε (c) modes. (d) Net correlation between IVS activity and different vibrational mode characteristics in AgSePh, evaluated for the 9 identified modes between 0 and 140 cm$^{-1}$. The *x*, *y*, and *z* directions correspond to the *a*, *b*, and 5.6-degrees-off-*c* crystallographic directions, respectively. See also vibrational mode animations included with the Online Supplementary Material.

The four primary IVS modes (α, β, γ, and δ) have corresponding peaks in the NRRS spectrum at 29.0, 61.2, 99.6, and 105.5 cm$^{-1}$, respectively (Table 1). The observation of these same four vibrational modes associated with exciton-phonon coupling by both techniques indicates that they couple to both the ground electronic state (as shown by NRRS) and the excited electronic state (as revealed with resonant IVS).

Due to the large unit cell of AgSePh consisting of 52 atoms, it is expected that many more Raman-active modes should exist in this material beyond the four modes identified by IVS. Indeed, the NRRS spectrum reveals >10 resolvable peaks in the spectral range from 0 cm$^{-1}$ to 140 cm$^{-1}$ (Figure 3a). As these vibrations are not observed in the IVS spectrum, we conclude that they correspond to Raman-active modes that do not couple strongly to the three primary excitonic transitions in AgSePh.

Another noteworthy difference between the results of the two techniques is the broadened linewidths in IVS compared to those seen in NRRS. First, we note that the linewidths are homogeneous in origin – rather than derived from sample inhomogeneity – due to the Lorentzian (rather than Gaussian) lineshapes seen in Figures 2d and S9.[49] Typically, homogeneous lifetime broadening due to phonon-phonon scattering is the main contribution to linewidth broadening in Raman spectroscopy, including NRRS and IVS. In addition to this contribution, the IVS linewidth is further broadened by phonon-exciton scattering[50–52] resulting from resonant excitation of the sample. Following photoexcitation at 5 K, excitons in the two higher-energy states ($X_2$ and $X_3$) scatter into the lowest energy state ($X_1$) on a ~0.5 ps timescale, followed by a slower ~3 ps depopulation timescale for the lowest-energy state (Figure S4). These electronic lifetimes are on the same order of magnitude as the coherent vibrational lifetimes extracted from time-domain fitting of the IVS data using Equation 1 (Table 1), indicating that linewidth broadening due to phonon-exciton scattering in AgSePh contributes significantly to linewidth broadening in resonant IVS – but not in non-resonant Raman scattering (NRRS). These fast exciton-phonon scattering rates may also contribute to the low photoluminescence quantum yield (~2%) of AgSePh at 5 K[37].

**Mapping observed frequencies onto atomic displacements in AgSePh**

To obtain a physical interpretation of the vibrational modes identified with IVS and NRRS, we compare the experimental vibrational spectroscopy results to lattice dynamics calculations performed using density functional perturbation theory (DFPT) approaches. A detailed description of how these calculations were performed can be found in the Methods section.

The calculated Raman spectrum and phonon density of states (PHDOS) are plotted alongside the experimental IVS and NRRS spectra in Figure 3a. As with the NRRS data, multiple vibrational modes were found in the calculated Raman spectrum, a subset of which are IVS active. We note that the peaks of interest are in the low frequency region (<140 cm$^{-1}$) and only separated by ~10

cm$^{-1}$ in several instances, whereas differences between calculated spectra and experimentally measured frequencies can be on the order of tens of wavenumbers[53] using first-principle approaches such as DFPT. To better match the peak structure observed in the experimental data set, the calculated spectrum was red-shifted by 8 cm$^{-1}$ for frequencies above 37 cm$^{-1}$. Peak assignments between the IVS, NRRS, and calculated Raman spectra were made based on nearest frequency modes and an assumption of correspondence between the data sets, as listed in Table 1 for the IVS active modes (see also Figure S10). A complete listing of all 9 vibrational modes identified in this study over the region of interest (0-140 cm$^{-1}$) and their frequencies can be found in Table S1. A discussion on the possible sources of error in the computed PHDOS and Raman spectra can be found in Supplementary Note 5.

Using DFPT, the atomic displacements of vibrational modes in AgSePh can be visualized in real-space. Mode α, which has a clear IVS signal, is shown in Figure 3b. The motion primarily consists of silver atom motion in the *z*-direction alternating between positive and negative displacements for adjacent silver atoms. (We define the *x*, *y*, and *z* directions as corresponding to the *a*, *b*, and 5.6-degrees-off-*c* crystallographic directions). In contrast, Mode ε (Figure 3c), which presents strongly in the calculated Raman spectrum but is not observed *via* IVS, consists primarily of phenyl ring twisting motion in the organic layer, with limited motion of the selenium atoms and almost stationary silver atoms.

To more systematically understand the modeled vibrational displacements and how they contribute to exciton-phonon coupling, we analyzed the simulated atomic motions to look for correlations with IVS activity. First, we considered the relative magnitude of atomic displacements, Cartesian-projected component motions, and changes in bond lengths. This analysis was performed on the five IVS-active and four IVS-inactive modes that could be clearly identified and assigned in our data set. Relative contributions of selected atomic motions to the key vibrational modes are presented in Table 1; a complete list of motional contributions to all 9 of the mapped vibrational modes can be found in Tables S2 and S3.

**Table 1.** Summary of experimental and simulation results describing key vibrational modes in AgSePh. See also Tables S1, S2, and S3.

| Mode Designation | NRRS at 78 K (cm$^{-1}$) | IVS at 5 K (cm$^{-1}$) | Calculated Raman (cm$^{-1}$) | IVS Lifetime at 5 K (ps) | Ag-Se Displacement Contribution[a] | Ag $z$-Motion Contribution[b] |
|---|---|---|---|---|---|---|
| Zeta (ζ)  | 25.5  | 24.1  | 23.7  | -   | 0.33 | 0.57 |
| Alpha (α) | 29.0  | 29.3  | 27.1  | 4.6 | 0.10 | 0.35 |
| Beta (β)  | 61.2  | 62.1  | 60.2  | 4.7 | 0.41 | 0.06 |
| Epsilon (ε) | 91.5 | -  | 94.7  | -   | 0.02 | 0.03 |
| Gamma (γ) | 99.6  | 98.7  | 102.4 | 2.4 | 0.15 | 0.07 |
| Delta (δ) | 105.5 | 103.7 | 106.4 | 3.4 | 0.11 | 0.09 |

*[a]Column value indicates atomic mass-weighted contribution of the specified displacement to total atomic displacements involved in the mode. All weighted displacements add to a value of 1.0.*
*[b]Column value indicates atomic mass-weighted contribution of the specified directional motion to the total of all atomic directional motions. All weighted directional motions add to a value of 1.0.*

Next, the motional contributions to all modes identified using DFPT were compared to generate a correlation score predicting which atomic motions are most strongly correlated with IVS activity. The correlation score is effectively a predictor of exciton-phonon coupling in AgSePh, estimating the degree to which certain atomic displacements are likely to impact the electronic structure of the material. A high correlation score indicates that a particular atomic displacement is likely to be important in exciton-phonon coupling. Figure 3d shows the correlation scores for each characteristic we evaluated, with more positive scores indicating stronger correlation. Full details on the method for calculating and assigning correlation scores can be found in Supplementary Note 4.

The strongest correlation with IVS signal is found for Ag motion in the $z$-direction. In Figure 1d, we established that the band edge electron densities are contained predominantly within the inorganic 2D plane, making it intuitive that Ag displacements in the $z$-direction (out of the 2D plane) are likely to disrupt the electronic structure of the inorganic layer. The second strongest correlating motion is the change in Ag-Se interatomic bond spacing, which is similarly central to the valence electron density (Figure 1d). Overall, we note the similarity of behavior observed here in AgSePh to exciton-phonon coupling in other 2D semiconductors. Specifically, out-of-plane homopolar phonons and in-plane polar optical phonons have been identified as the most significant contributors to exciton scattering in other 2D semiconductors,[50,54,55] agreeing with identification here of Ag motion in the *z*-direction and Ag-Se interatomic bond spacing, respectively, as important displacements. Additionally, these results corroborate a recent finding from Schriber *et al.*[32], who conclude that bonding among the 2D network of Ag atoms is important for light emission in AgSePh and in two related compounds, AgSPh and AgTePh.

Returning to the modes previously introduced in Figure 3b and Figure 3c, we see in Table 1 that Mode α exhibits a much larger Ag *z*-directed atomic displacement contribution than Mode ε, helping to explain why there is a greater degree of exciton-phonon coupling in Mode α than Mode ε, as measured with IVS.

**Vibrational modes strongly influencing light emission**

To identify which modes strongly influence the luminescence properties of AgSePh, we performed temperature-dependent photoluminescence (PL) spectroscopy on AgSePh crystals. Figure 4a shows a map of intensity-normalized PL spectra of AgSePh from 5 to 300 K in 5 K increments. With increasing temperature, the PL initially blue-shifted from 5 to 30 K before reversing direction and red-shifting from 30 K to 300 K. The shift was smooth without an abrupt change in position, suggesting an absence of phase transitions in this temperature range. Concurrent with the gradual peak shift, the PL linewidth monotonically broadened as a function of temperature, as expected for a lineshape influenced by exciton-phonon coupling effects.

At very low temperature, the emission spectrum of AgSePh splits into multiple peaks and as many as four distinct optical transitions could be resolved at 5 K (Figure 4b). Fitting of these four peaks to Voigt functions results in peak energies of 2.6928, 2.7087, 2.7223, and 2.7367 eV, yielding peak energy separations of 15.9, 13.6, and 14.5 meV (128, 110, and 117 cm$^{-1}$), respectively. The roughly constant separation in energy suggests the origin of peak splitting to be phonon sidebands[56] arising from strong interaction between an exciton and a dominant phonon having an energy of ~13-16 meV (~105-129 cm$^{-1}$).

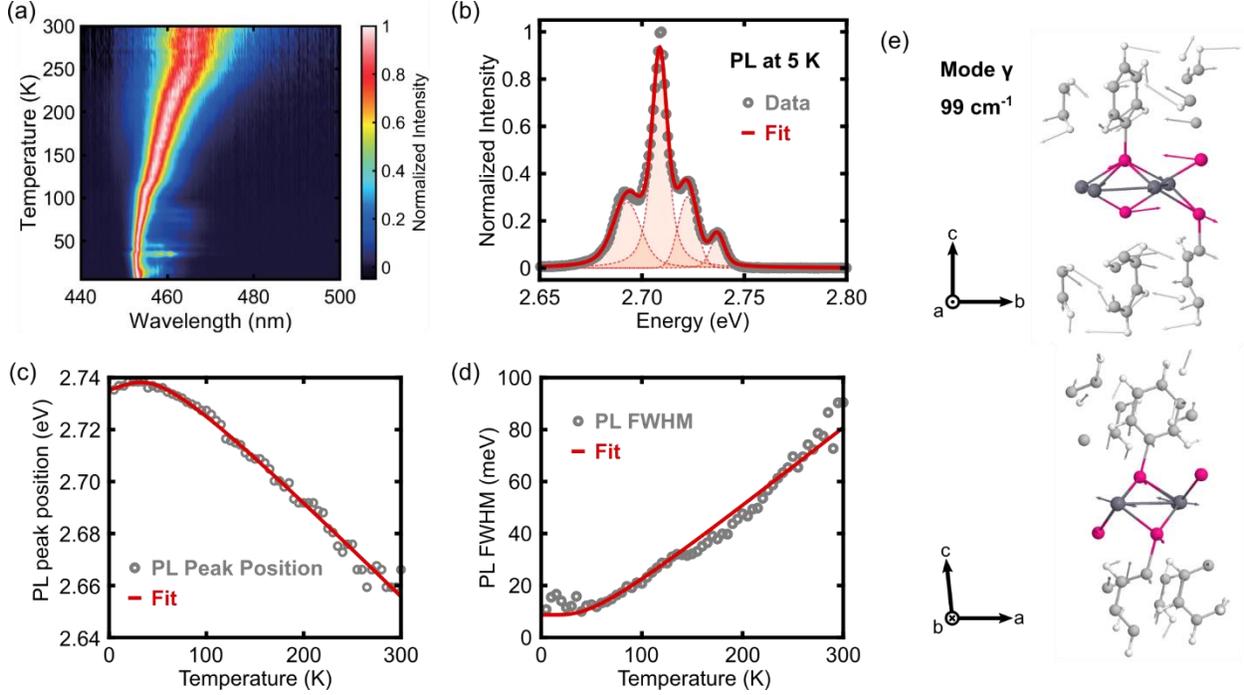

**Figure 4. Exciton-phonon coupling affecting light emission in AgSePh.** (a) Intensity-normalized temperature-dependent photoluminescence spectra from 5 to 300 K in 5 K steps. (b) Photoluminescence spectrum at 5 K showing peak splitting and their fits to Voigt functions centered at 2.6928, 2.7087, 2.7223, and 2.7367 eV. (c) Temperature-dependent photoluminescence peak shift and corresponding fit to Equation 2. (d) Temperature-dependent photoluminescence linewidth broadening and corresponding fit to Equation 3. (e) DFPT-simulated lattice displacements of mode γ (98.7 cm$^{-1}$ or 12.2 meV) viewed along two different in-plane axes. See also vibrational mode animations included with the Online Supplementary Material.

An analysis of the temperature-dependent PL peak shift can be used to extract the energy of the vibration most strongly coupled to the excitonic state responsible for light emission. Assuming a quasi-harmonic oscillation and a temperature-independent exciton binding energy, the temperature dependent electronic band gap (*i.e.* quasiparticle gap) of a semiconductor at constant pressure, $E_g(T)$, can be described by[57,58]

$$E_g(T) = E_0 + A_{TE}T + A_{EP}\left(\frac{2}{\exp\left(\frac{E_{ph}}{k_B T}\right) - 1} + 1\right), \quad (2)$$

where $E_0$ is an unrenormalized band gap, $A_{TE}$ determines the weight of thermal expansion interaction, $A_{EP}$ accounts for the strength of exciton-phonon interaction, $E_{ph}$ is the energy of a dominant phonon, and $k_B$ is the Boltzmann constant. Assuming a temperature-independent exciton binding energy, a fit of Equation 2 to the temperature-dependent peak position of AgSePh (Figure 4c) yields an $E_0$ of $2.770 \pm 0.004$ eV, $A_{TE}$ of $0.12 \pm 0.11$ meV/K, $A_{EP}$ of $-35.2 \pm 3.3$ meV, and $E_{ph}$ of $12.3 \pm 2.8$ meV.

To further confirm the fitted phonon energy, $E_{ph}$, we applied the value extracted from the peak-shift analysis (Figure 4c) to an analysis of the temperature-dependent PL linewidth broadening of AgSePh. Figure 4d compares the PL full width at half maximum (FWHM) as a function of temperature to the predicted broadening. We found that the PL FWHM was well-fitted by a model comprised of a temperature-independent linewidth, $\Gamma_0$, and a contribution from a single vibrational mode with an energy of 12.3 meV[59–61]:

$$\Gamma_{homo}(T) = \Gamma_0 + \frac{\Gamma_{ph}}{\left[\exp\left(\frac{E_{ph}}{k_B T}\right) - 1\right]}. \quad (3)$$

Here, $\Gamma_{ph}$ is an exciton-phonon coupling strength parameter, and $\Gamma_0$ and $\Gamma_{ph}$ were found to be $9 \pm 1$ meV and $44.0 \pm 1.0$ meV, respectively.

Overall, these findings indicate that a single dominant vibrational mode having energy of ~12-16 meV (~97-129 cm$^{-1}$) is responsible for the low-temperature PL spectral splitting, the temperature-dependent peak shifting, and the temperature-dependent linewidth broadening in AgSePh. Further, we see agreement between the exciton-phonon coupling behavior determined from PL analysis and the excitonic state-dependent IVS vibrational spectrum in Figure 2e, which showed that the γ mode at 99 cm$^{-1}$ dominates coherent lattice dynamics of the X$_1$ state.

The X$_1$ state is the lowest-energy excitonic state in AgSePh and also the state responsible for light-emission[35,37]. In Figure 2e we observed that the β mode at 62 cm$^{-1}$ was the dominant mode for the highest energy excitonic state (X$_3$) based on IVS signal intensity, while the γ mode at 99 cm$^{-1}$ was the highest intensity peak for the lowest-lying excitonic state (X$_1$). The relative coupling strength of the γ mode may also be underestimated because the intensity of an IVS peak is expected to vary inversely with the square of vibrational frequency ($I \propto 1/\omega^2$)[17,62].

Figure 4e shows the calculated vibrational motion of the γ mode viewed from two orthogonal in-plane directions. Overall, the γ mode is a strongly hybridized mode involving the wagging of adjacent phenyl rings and corresponding stretching of the Ag-Se bonds, which we identified earlier to be correlated with strong exciton-phonon interaction.

**Temperature-dependent vibrational dynamics**

Impulsive vibrational spectroscopy (IVS) was performed as a function of sample temperature from 5 K to 200 K using the same experimental setup and conditions discussed previously. The time-domain vibrational dynamics as a function of temperature are displayed in Figure 5a. We focus here on the blue edge of the highest energy excitonic peak ($X_3$) because it has the highest signal-to-noise ratio across the widest temperature range (specific wavelength ranges and other details are included in Supplementary Note 2). At each temperature, the time-domain vibrational dynamics were fit to the sum of four exponentially decaying sine waves. Fits to the data and a summary of the fitting parameters for all temperatures can be found in Figure S14 and Table S4, respectively.

Vibrational dephasing times for the four dominant IVS modes are plotted in Figure 5b as a function of sample temperature. Overall, dephasing becomes faster at higher temperatures, as expected due to increasing population of the incoherent phonon bath with increasing temperature.[63]

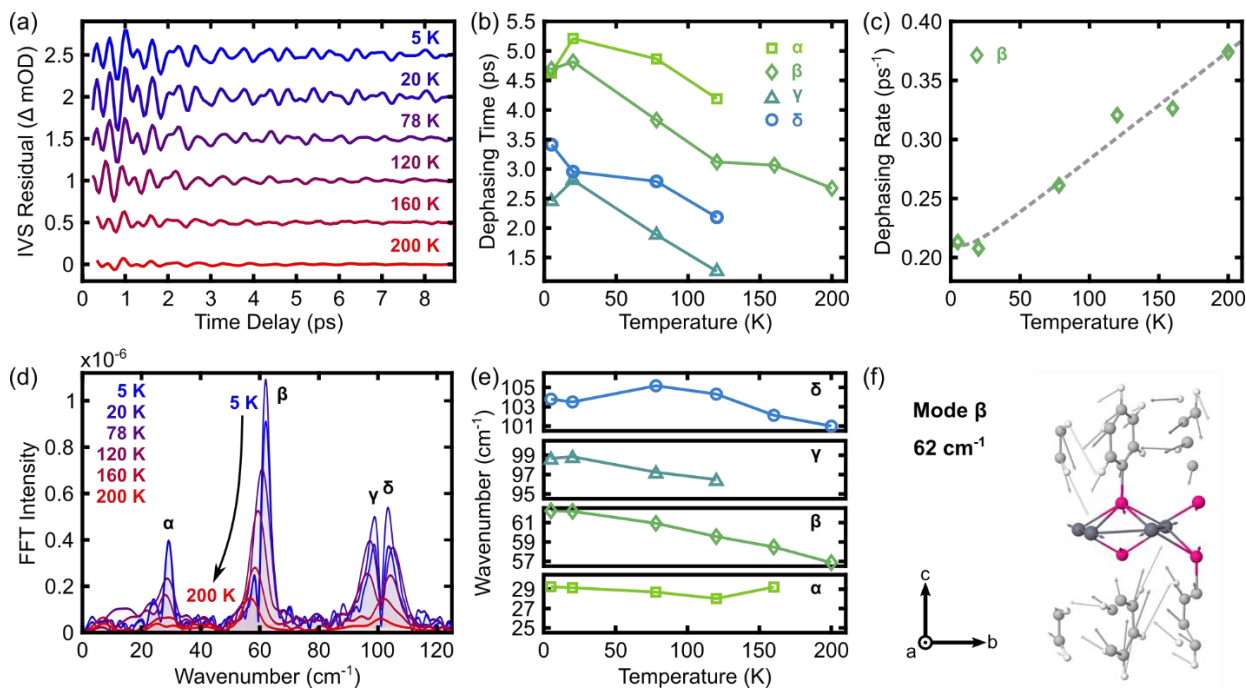

**Figure 5. Temperature-dependent vibrational dynamics.** (a) Time-domain IVS coherent vibrational response from 5 K to 200 K. (b) Temperature-dependent dephasing times for the four primary IVS-active modes, determined by fitting the data in (a) to the sum of four decaying sine waves. (c) Dephasing rate of the β mode as a function of temperature, fitted with a cubic overtone model. (d) Temperature-dependent IVS spectrum obtained by Fourier transform of the curves in (a). See also Figure S13. (e) Dependence of IVS mode frequency on sample temperature. Note that some peaks become irresolvable at higher temperatures, leading to series truncation. (f) Real-space displacements of the β vibrational mode, showing shearing motion of adjacent phenyl rings. See also vibrational mode animations included with the Online Supplementary Material.

Temperature-dependent dephasing rates can be fit to a cubic overtone model, in which one coherent optical phonon decays into two acoustic phonons of equal energy and opposite wave vectors.[63,64] This model provides a simplified picture of vibrational dephasing, but yields a useful measure of anharmonic coupling strength[63] and the relative magnitude of different sources of dephasing. We focus on the 62 cm$^{-1}$ mode as it is the best resolved feature in the IVS spectrum, especially at higher temperatures. The dephasing rate (or the inverse of the dephasing lifetime) is fitted to the cubic overtone model using

$$\Gamma = \Gamma_0 + \gamma_0 \left[ 1 + \frac{2}{\exp\left(\frac{\hbar\omega}{2k_B T}\right) - 1} \right], \qquad (4)$$

where $\Gamma_0$ is a temperature-independent scattering rate (*i.e.* defect scattering), $\gamma_0$ is the anharmonic coefficient, and $\omega$ is the vibrational mode frequency. As shown in Figure 5c, this model successfully captures the observed trend in dephasing rate with fitting parameters of $\Gamma_0 = 0.190 \pm 0.025$ ps$^{-1}$ and $\gamma_0 = 2.1\times10^{-2} \pm 0.4\times10^{-2}$ ps$^{-1}$, suggesting that both temperature-dependent and temperature-independent sources of scattering are significant. Again, we note that the exciton lifetime is <10 ps at all temperatures, and on the same order of magnitude as the vibrational coherence time (see Figure S12). Coherent vibrations on the electronic excited potential energy surface would likely be scattered by an electronic transition back to the ground state, limiting the conclusions that can be reached with a phonons-only analysis.

Temperature-dependent IVS spectra (obtained by Fourier Transform of the time-domain data shown in Figure 5a) are displayed in Figure 5d. These curves show notable shifts in peak width and position as a function of temperature, and the magnitude of these shifts varies among the different vibrational modes. Peak broadening results directly from homogeneous lifetime broadening, as discussed in the preceding paragraph. Additionally, we observe a temperature-dependent frequency shift of all modes (Figure 5e), revealing the role of mode anharmonicity[65].

Of the modes appearing in the IVS spectrum, the β mode exhibits the largest temperature-dependent frequency red-shift ($\Delta\omega/\omega$ = 9% from 5 K to 200 K). Atomic displacements of the β mode are illustrated in Figure 5f. This mode is dominated by the phenyl rings, which slide past one another as they undergo collective wagging motion. In contrast, the α mode (Figure 3b) – which showed the smallest temperature-dependent frequency shift ($\Delta\omega/\omega$ = <1% from 5 K to 160 K) – consists primarily of Ag atomic motion in the *x*- and *z*-directions, with minimal contribution of the phenyl rings. These findings suggest that the covalently-bound silver atoms within the inorganic 2D plane exist in a more harmonic displacement potential than the phenyl rings, which interact with adjacent rings *via* co-facial van der Waals attractive forces and steric repulsive forces, leading to a greater degree of anharmonicity for the β mode.

When compared to other material systems, these findings clearly reveal the truly hybrid nature of layered 2D AgSePh. Over a 5 K to 200 K temperature range, the anharmonic frequency shift of the dominant modes in AgSePh (~0-9%) is much larger on average than purely inorganic crystalline systems (~1-2%)[66–69] but smaller than crystalline polycyclic aromatic organic molecules like anthracene and naphthalene (~7-20%)[70]. Further, the β mode—which most heavily involves motion of the organic phenyl moiety—has an anharmonic frequency shift in the range reported for the organic molecules, while the α mode—primarily confined to the inorganic sublayer—shows a frequency shift more in-line with the fully inorganic systems.

## Conclusions

Overall, our results provide new insight into exciton-phonon interactions in AgSePh and, more broadly, in low-dimensional hybrid organic-inorganic semiconductors. The unique electronic structure of AgSePh, showing three distinct excitonic features, allows for varying degrees of exciton-phonon coupling among different excitonic states and the vibrational modes of the system. The combination of IVS, PL, and DFPT techniques enabled identification of the γ mode – characterized by the wagging of adjacent phenyl rings and corresponding stretching of the inorganic Ag-Se bonds – as the mode most strongly coupled to the emissive $X_1$ state. Moreover, temperature-dependent IVS revealed increased mode anharmonicity in hybrid organic-inorganic AgSePh compared to all-inorganic crystals, especially for vibrational modes that involve significant motion of the phenyl ring (such as the β mode). Finally, these experiments highlight the detailed information that can be obtained by integrated time-domain, frequency-domain, and computational efforts to understand exciton-phonon interactions in hybrid nanomaterials.

These findings also suggest the ability to tune optoelectronic properties of metal organochalcogenolates through synthetic modification like chalcogen substitution[32,35,37] or functionalization of the organic layer[31,34,71]. For example, modifying the phenyl rings to create steric hindrance or allow for interlayer cross-linking may increase the rigidity of the structure, leading to higher photoluminescence efficiency[72]. Furthermore, it may be possible to induce or suppress exciton-phonon interactions to realize targeted properties like minimized broadband emission from self-trapped excitons[36,37,73] or improved charge-carrier screening and transport properties *via* polaron formation. Finally, engineering excitonic behavior through molecular doping of the organic layer may offer an orthogonal route to achieve desired performance[74,75].

## Experimental procedures

### Resource availability

*Lead contact*

Further information and requests for resources, data, and materials should be directed to and will be fulfilled by the lead contact, William A. Tisdale (tisdale@mit.edu).

*Materials, data, and code availability*

All materials, data, and code generated and used in this study are available upon sensible requests to the lead contact.

**Synthesis of AgSePh thin films.** Thin films of silver phenylselenolate (AgSePh) for ultrafast spectroscopy were made by chemical transformation of metallic silver with an organic diselenide vapor in the presence of a solvent vapor[26,33]. Metallic silver (Ag) films with thickness of 10 nm were deposited on single-crystal quartz substrates using a thermal evaporator. Each silver film was sealed inside a microwave reaction vial along with two culture tubes separately containing ~30 mg diphenyl diselenide (Ph$_2$Se$_2$) and 200 µL dimethylsulfoxide (DMSO). The sealed reaction vessels were heated in an oven at 100 °C for 5 days, yielding yellow films of AgSePh.

**Synthesis of AgSePh crystals.** Large (~millimeter) AgSePh crystals for Raman and PL spectroscopy were synthesized by an organic single-phase reaction[31] by mixing 5 mL of 10 mM AgNO$_3$ solution in propylamine (PrNH$_2$) with 5 mL of 10 mM Ph$_2$Se$_2$ solution in toluene and holding at room temperature for 5 days.

**Optical absorption spectroscopy.** Optical absorption measurements were performed on AgSePh thin films in transmission mode using a Cary 5000 UV-Vis-NIR spectrophotometer, and on ground crystal powders using the same spectrophotometer with a PIKE Technologies DiffuseIR accessory. For the measurements on ground crystals, solid samples were prepared by grinding with dry potassium bromide (KBr) to a ~1 wt% dilution, and diffuse reflectance spectra were normalized to a 100% KBr baseline. The obtained diffuse reflectance spectra were converted into absorption spectra using the Kubelka-Munk transform[76], $F(R) = \frac{(1-R)^2}{2R}$, where $F(R)$ is the Kubelka-Munk function having a value proportional to the sample's absorption coefficient, and $R$ is the relative reflectance of the sample with respect to the 100% KBr baseline.

**Photoluminescence spectroscopy.** Photoluminescence measurements were performed on AgSePh crystals using an inverted microscope (Nikon, Ti-U Eclipse). Samples were mounted on a cold finger in a microscopy cryostat (Janis Research, ST-500) and cooled by flowing liquid helium. A 405 nm laser diode (Picoquant, LDHDC-405M, continuous wave mode) was used for photoexcitation, and the emission spectrum was imaged through a spectrograph (Princeton Instruments, SP-2500) mounted with a cooled charge-coupled detector (Princeton Instruments, Pixis).

**Non-resonant Raman scattering (NRRS).** Non-resonant Raman spectra were collected in a backscattered geometry on the same microscopy setup used for photoluminescence measurements. A 785 nm narrow-band continuous-wave laser (Ondax) was used as an excitation source and the scattered laser light was passed through a set of volume holographic grating notch filters (Ondax) before being directed into the spectrograph mounted with the cooled charge-coupled detector. The role of the notch filters is to suppress the Rayleigh scattering, allowing an accurate measurement of vibrational modes to <10 cm$^{-1}$. The resolution of the system is approximately 0.5 cm$^{-1}$.

**Resonant impulsive vibrational spectroscopy (IVS).** A 375 nm wavelength pump pulse at 200 kHz repetition was used to excite the sample. Unless otherwise noted, the pump power was 1100 µW, which corresponds to 5.5 µJ/pulse. A broadband laser pulse covering at least the 405-485 nm spectral region with a power of <100 µW was used to probe the sample after a variable time delay from the excitation. A probe spot size of approximately 7900 µm$^2$ was used with the pump spot being slightly larger, leading to an excitation density of approximately 3.0×10$^5$ photons/µm$^2$/pulse. The pump pulse duration was 71.2±0.8 femtoseconds (fs) full width at half maximum (FWHM) as measured by frequency resolved optical gating (FROG) utilizing the optical Kerr effect (OKE).[77] Further details on the experimental setup and pulse duration measurement can be found in Supplementary Note 1.

**Density Functional Theory calculations.** Density functional theory (DFT), phonon density of states (PHDOS), and Raman spectra calculations were performed using the VASP planewave pseudopotential code[78–81]. A full structural relaxation of the experimentally determined 52-atom primitive unit cell with an ionic force convergence threshold of $1.0 \times 10^{-7}$ $eV/Å$ was performed prior to the phonon calculations. For the structural relaxation, an $8 \times 8 \times 1$ Monkhorst-Pack k-point grid with a Gaussian smearing of 0.01 eV was used. The valence electron – ionic core interactions were treated used projector augmented wave (PAW) potentials [82,83]. A planewave kinetic cutoff energy of 500 eV was used. The Perdew–Burke-Ernzerhof (PBE) generalized gradient approximation to the exchange correlation functional[84] was used in all calculations. Taking the DFT-relaxed equilibrium structure, the Phonopy code[85] was used to generate $3 \times 3 \times 1$ supercell structures with atomic displacements in order to calculate the PHDOS over the Brillouin zone. These 152 structures were then used to compute the force-constant Hessian matrix *via* the density functional perturbation theory (DFPT) capabilities in VASP[86,87]. The Phonopy code was used to post-process the calculation data, generating the simulated total PHDOS spectra, the Cartesian-direction projected PHDOS spectra, the atom-projected PHDOS spectra, and the *xyz* trajectories of the phonon eigenmodes. The Phonopy-Spectroscopy code[88] was used in conjunction with VASP dielectric constant calculations[89–91] for the displaced structures in order to calculate the Raman intensities and simulated Raman spectrum of the vibrational modes. Visualizations of the vibrational modes were obtained using the VMD software[92]. As another point of comparison for the VASP/Phonopy calculated spectra, first principles molecular dynamics simulations were performed using the Qbox code[93], from which vibrational power spectra were calculated and found to have generally good agreement.

## Data availability

The authors declare that the data supporting the findings of this study are available within the paper and its Supplementary Information file. All other relevant data supporting the findings of this study are available on request.

## Acknowledgements

Materials synthesis and characterization was supported by the U.S. Army Research Office under Award Number W911NF-23-1-0229. Ultrafast spectroscopy experiments were funded by the U.S. Department of Energy, Office of Science, Basic Energy Sciences under Award Number DE-SC0019345. D.C.Y. was supported through an appointment to the Intelligence Community Postdoctoral Research Fellowship Program at Massachusetts Institute of Technology, administered by the Oak Ridge Institute for Science and Education through an interagency agreement between the U.S. Department of Energy and the Office of the Director of National Intelligence. This research used resources of the National Energy Research Scientific Computing Center (NERSC), a U.S. Department of Energy Office of Science User Facility operated under Contract No. DE-AC02-05CH11231. E.R.P. was supported by the US Department of Defense through the National Defense Science & Engineering Graduate (NDSEG) Fellowship Program. W.S.L was partially supported by the Seoul Broadcasting System Foundation Overseas Doctoral Program Scholarship.

## Author contributions

W.P. and W.S.L. synthesized the material. E.R.P. performed impulsive vibrational spectroscopy and data analysis. W.P. performed photoluminescence spectroscopy and data analysis. W.P. and W.S.L. performed non-resonant Raman spectroscopy. D.C.Y. and J.C.G. performed and supervised density functional theory calculations, respectively. E.R.P., W.P., and W.A.T. wrote the manuscript with input from all authors. All authors reviewed the manuscript.

## ORCID

Eric R. Powers:              0000-0003-1342-5801

Watcharaphol Paritmongkol:   0000-0003-1638-6828

Dillon C. Yost:              0000-0003-0854-1218

Woo Seok Lee:                0000-0001-9188-5104

Jeffrey C. Grossman:         0000-0003-1281-2359

William A. Tisdale:          0000-0002-6615-5342

## Declaration of interests

The authors declare no competing interests.

## Inclusion and diversity

We support inclusive, diverse, and equitable conduct of research.

## Additional information

**Supplementary information** The online version contains supplementary material available at https://doi.org/XXXXXX/XX.

**Correspondence** and requests for materials should be addressed to:

William A. Tisdale – Department of Chemical Engineering, Massachusetts Institute of Technology, Cambridge, Massachusetts 02139, United States; orcid.org/0000-0002-6615-5342; E-mail: tisdale@mit.edu

Supplementary Information for:

# Coherent Exciton-Lattice Dynamics in a 2D Metal Organochalcogenolate Semiconductor

Eric R. Powers[1,4], Watcharaphol Paritmongkol[1,2,4], Dillon C. Yost[3], Woo Seok Lee[1,3], Jeffrey C. Grossman[3], William A. Tisdale[1]*

[1]Department of Chemical Engineering, Massachusetts Institute of Technology, Cambridge, Massachusetts 02139, United States

[2]Department of Chemistry, Massachusetts Institute of Technology, Cambridge, Massachusetts 02139, United States

[3]Department of Materials Science, Massachusetts Institute of Technology, Cambridge, Massachusetts 02139, United States

[4]These authors contributed equally.

*Correspondence: tisdale@mit.edu

**Current affiliations**:

W.P. Department of Materials Science and Engineering, School of Molecular Science and Engineering, Vidyasirimedhi Institute of Science and Technology (VISTEC), Rayong 21210, Thailand

D.C.Y. Design Physics Division, Lawrence Livermore National Laboratory

## Contents

## Section A: Structural Characterization

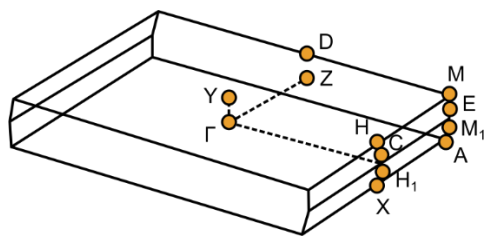

**Figure S1. Brillouin zone image of AgSePh adopting $P2_1/c$ space group.**

# Section B: IVS Experimental Methods Discussion and Additional Figures

**Supplementary Note 1: Impulsive Vibrational Spectroscopy Setup**

Impulsive vibrational spectroscopy (IVS) was performed using a Spirit 1040-8 ultrafast laser (Spectra-Physics) operating at 200 kHz. A portion of the 1040 nm fundamental light was directed to a commercial two-stage non-collinear optical parametric amplifier (NOPA) with a second harmonic stage (Spirit-NOPA, Spectra-Physics), where it was converted to 375 nm pump light. The NOPA output was filtered using a 375 nm bandpass filter (Chroma AT375/28x), compressed using two fused silica prisms to minimize pulse duration, and modulated to an intensity of 1.1 mW using neutral density filters.

A second portion of the 1040 nm light was frequency-doubled to 520 nm in a β-barium borate (BBO) crystal. The residual 1040 nm was filtered out using a dichroic beam splitter, and the remaining ~150 mW 520 nm pulse was then focused into a 3 mm sapphire window with a 100 mm focal length lens to generate a broadband supercontinuum. The resultant white light was recollimated and filtered using either a 500 nm shortpass filter (Thorlabs) or two consecutive filters (Thorlabs notch filter 533nm, 17nm full width at half maximum FWHM; PIXELTEQ magenta color filter 544 nm, 124 nm FWHM) to create the probe pulse covering a minimum spectral range of 405-485 nm.

The pump pulse was modulated at 5 kHz using a mechanical chopper (Thorlabs), and the time delay between the pump and probe pulses was controlled using a mechanical delay stage (Newport). The pulses were overlapped using a 200 mm focal length mirror to a spot size of ~7900 µm$^2$ at the sample, which was mounted in a tower cryostat under vacuum (Janis Research, ST-100). The transmitted probe light was recollimated and directed to a spectrometer and high-speed data acquisition system (Ultrafast Systems) with a 90 µs collection window, corresponding to ~18 consecutive probe pulses for each successive "pump on" and "pump off" acquisition period. The displayed data were collected from a minimum of 4 different locations in the sample to confirm sample homogeneity, and approximately 8-12 scans were taken at each location to rule out significant sample degradation during data collection. A chirp correction to the data was performed based on a cross correlation of the pump and probe pulses as measured in a 2 mm fused silica window. The photoluminescence background signal was also subtracted using an average of 5 data points from before time zero.

For pump pulse metrology, an autocorrelation was performed on the pump pulse using the optical Kerr effect in a 2 mm fused silica window. The temporal and spectral pulse profile was retrieved using the frequency-resolved optical-gating (FROG) method[1] with code adapted from the Trebino group[2] as illustrated in Figure S2 and Figure S3. The measured pump pulse duration of 71.2±0.8 fs FWHM sets a theoretical upper limit of ~468 cm$^{-1}$ for the highest frequency coherent vibration that can be measured using this IVS experimental setup, although the actual value is likely lower due to effects from the transform limit of the probe pulse[3,4] and the much higher signal-to-noise ratio needed when approaching the theoretical limit.

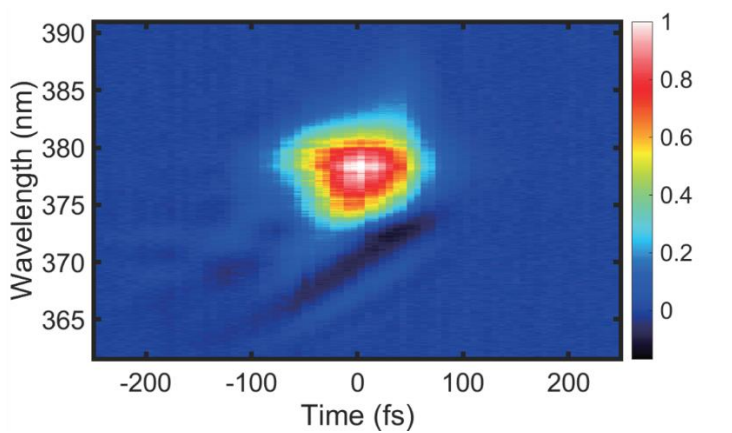

**Figure S2. Color map of pump autocorrelation results before deconvolution analysis.**

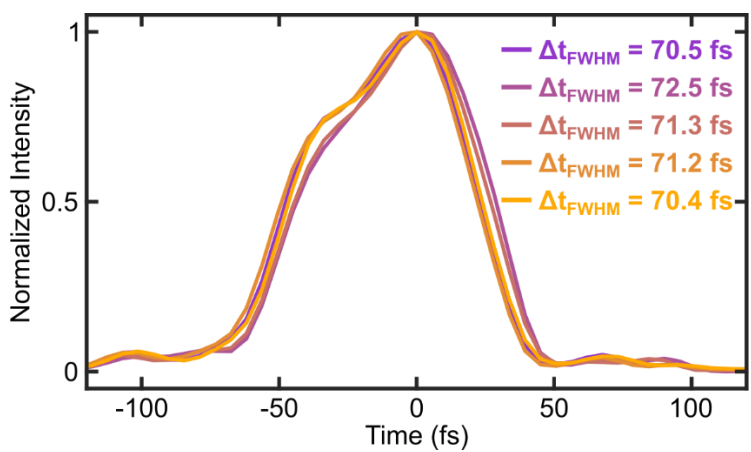

**Figure S3. Pump pulse temporal lineshape after execution of the FROG algorithm.** FROG analysis performed on the data shown in Figure S2. Different curves are the results from 5 separate runs of the FROG code using the same input data, validating the reproducibility of the algorithm.

**Supplementary Note 2: Impulsive Vibrational Spectroscopy Frequency Domain Analysis**

When fitting raw IVS data, the following procedure was followed. First, a wavelength region with a clear and in-phase oscillatory signal was identified in the 2D IVS data. As noted in the main text, the IVS signal is strongest near wavelengths at which the linear absorption spectrum has a high change in intensity with respect to wavelength[4]. Conversely, nodes in the IVS data and phase flips occur at the extrema of the absorption spectrum when the slope of the signal changes sign. For the complex electronic structure of AgSePh, this led to multiple nodes in the IVS signal. In the case of analyzing 5 K data, the IVS signal along the blue edge of the highest energy excitonic peak from 408 nm to 422 nm was summed to generate a one-dimensional curve of intensity versus time delay. For IVS data at temperatures above 5 K shown in Figure 5a and elsewhere, the following wavelength ranges were summed, always using the blue edge of the highest energy exciton — 20 K: 408-422 nm, 78 K: 413-424 nm, 120 K: 414-427 nm, 160 K: 414-429 nm, 200 K: 414-429 nm, and 300 K: 412-433 nm. A 20 ps time window was used to analyze the 5 K and 20 K data, with only a 10 ps window being necessary at 78 K and above due to faster IVS signal dephasing. After summing, the low frequency electronic dynamics were fitted and subtracted off to leave only the vibrational components. The fit was performed in MATLAB using a custom fitting equation. After trying various functional forms, the best fit to the electronic dynamics was found using the sum of 1) the tail of a Gaussian distribution, 2) a triple exponential decay, and 3) a constant offset, a total of 10 fitting parameters. The fit was determined by providing a reasonable initial guess and iteratively optimizing until an optimal stable solution was found.

Next, a Fourier transform was applied to convert the time-domain vibrational data to the frequency domain. A Kaiser-Bessel window function was applied to the time-domain data before the Fourier transform was performed[5], as this has been shown to reduce ringing effects in IVS data[4,6]. By an empirical evaluation, an alpha value of 1.0 was chosen to minimize ringing effects without excessively broadening the peak width. A total of 8000 additional zero-valued data points were appended to the end of the <400 time-domain data points to increase the spectral resolution (zero padding)[4,6]. The Fourier transform was performed using the fft function in MATLAB. The resultant data could be fitted to the sum of multiple Lorentzian peaks as shown in Figure 2d. A Lorentzian shape was found to provide a better fit than Gaussian and is consistent with homogeneous linewidth broadening mechanisms (see also Figure S7). It is worth noting that the apparent node between the γ and δ peaks at around 100 cm$^{-1}$ is an effect of performing a Fourier transform on two decaying sinusoidal waves near to each other in frequency, and we do not believe it to be physically meaningful. However, this effect led us to report all IVS frequencies based on the IVS experimental peak frequency, rather than using the center of the fitted Lorentzian due to difficulties fitting the data in this region. We also note that this analysis process involves summing the data across a wavelength region before Fourier transforming. Alternatively, the data can first be Fourier transformed at each wavelength and then summed, as has been reported by others[4]. This method was also tested and is shown in Figure S6, but it was found to generate a higher noise level than first summing the raw data due to random fluctuations in individual wavelength pixel intensities.

**Supplementary Note 3: Impulsive Vibrational Spectroscopy Time Domain Data Analysis**

The following procedure was used to fit the IVS vibrational data in the time domain. The functional form of the fit we employed is given by Equation 1 in the main text. The fit was performed iteratively, first optimizing for a single decaying sine term, then two terms, and so on. At each iteration, the initial guesses were based on the results of the previous iteration and the strongest peak frequencies identified in the corresponding frequency domain data. This process was continued until the fitting parameters remained constant between successive iterations and four decaying sine terms had been fitted, which all agreed well with the four strongest modes identified in the frequency domain. The fitting results at all temperatures are shown in Figure S14, and the fitted parameters are listed in Table S4. At 160 K and 200 K, it was only possible to fit three terms, which is consistent with the frequency domain results in Figure S13, where the γ peak intensity drops significantly at higher temperatures.

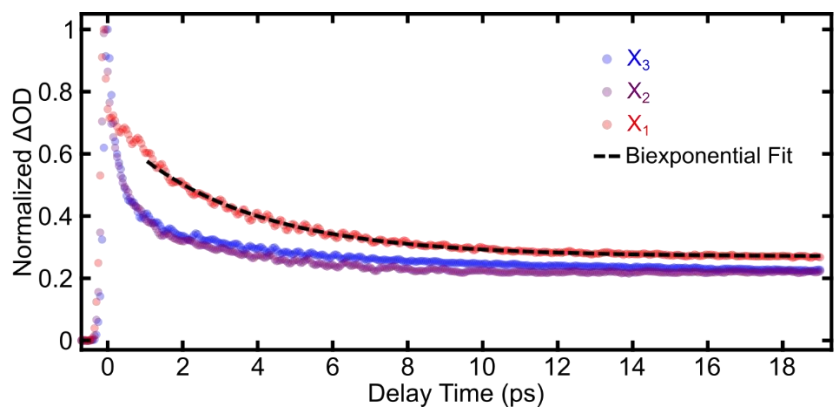

**Figure S4. Dynamics of the three bleach features in the AgSePh IVS spectrum at 5 K.** Dynamics for the first 0.5-1 ps consist primarily of decay from the higher energy $X_3$ (422-428 nm) and $X_2$ (440-446 nm) excitonic features to the lowest energy $X_1$ (451-457 nm) peak. The dynamics of the $X_1$ peak after 1 ps are fit with a biexponential decay, yielding a 3.4 ps fast lifetime, consistent with previously reported findings[7] and likely driven by exciton trapping or recombination through radiative and non-radiative pathways. The decay of the slow component was found to be effectively zero over the 20 ps time window, so the true long lifetime could be any value several hundred picoseconds or longer.

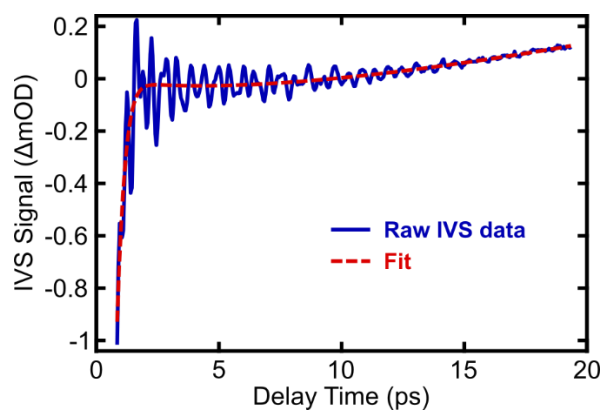

**Figure S5. Isolating coherent vibrational data from electronic dynamics.** The raw IVS data at 5 K extracted from Figure 2a is shown in blue. The fit to the data is shown in red, as discussed in Supplementary Note 2. Subtraction of the fit from the raw data yields the vibrational dynamics shown in Figure 2c.

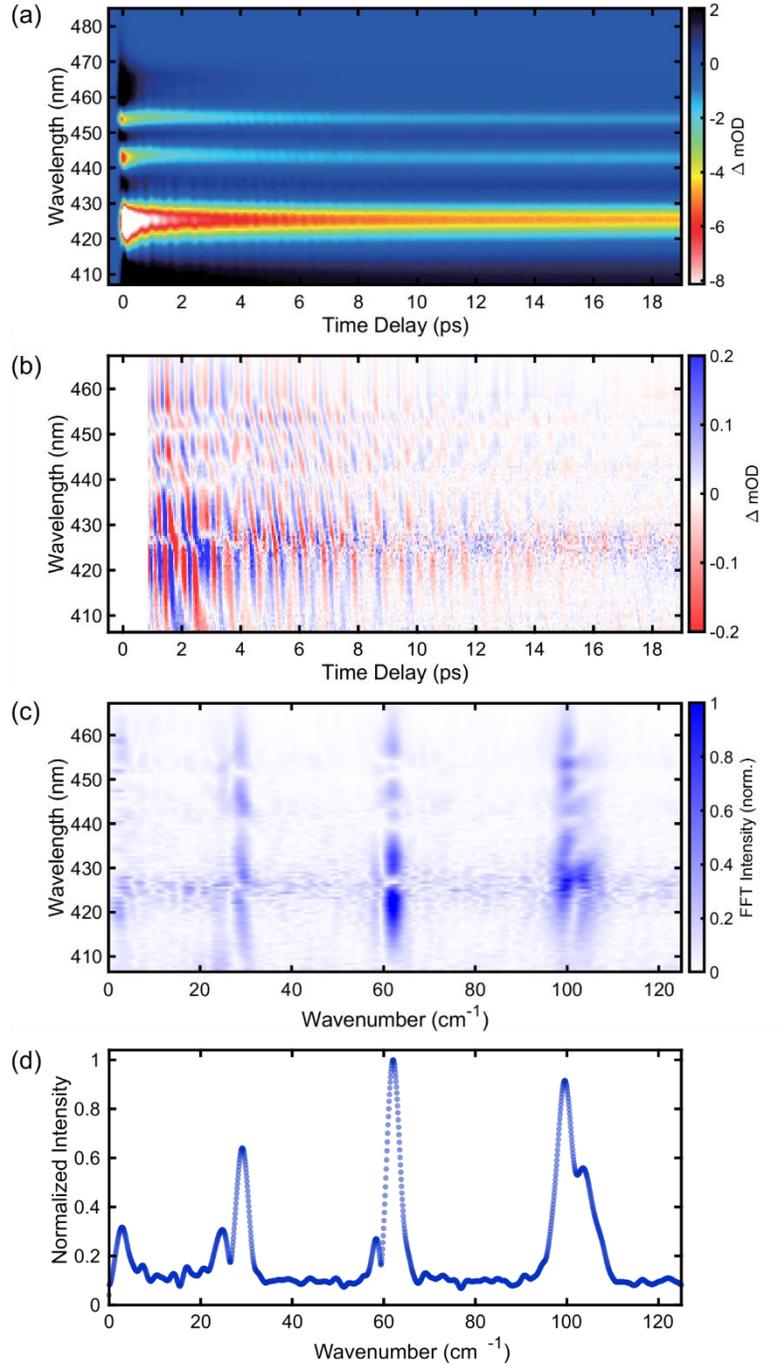

**Figure S6. Individual wavelength method of extracting IVS frequency domain data.** (a) IVS color map of AgSePh at 5 K using the same data set as analyzed in Figure 2 of the main text. (b) Coherent vibration residual data derived by individually subtracting off the electronic dynamics at each wavelength pixel rather than first summing the data and then performing the electronic dynamics subtraction only once. (c) Results of a Fourier transform to the data in panel (b), again performed at each wavelength pixel. These results show the same vibrational frequencies are present at all wavelengths but with varying intensities. (d) Sum of the 2D data in panel (c) across all wavelengths, resulting in a 1D IVS spectrum comparable to Figure 2d.

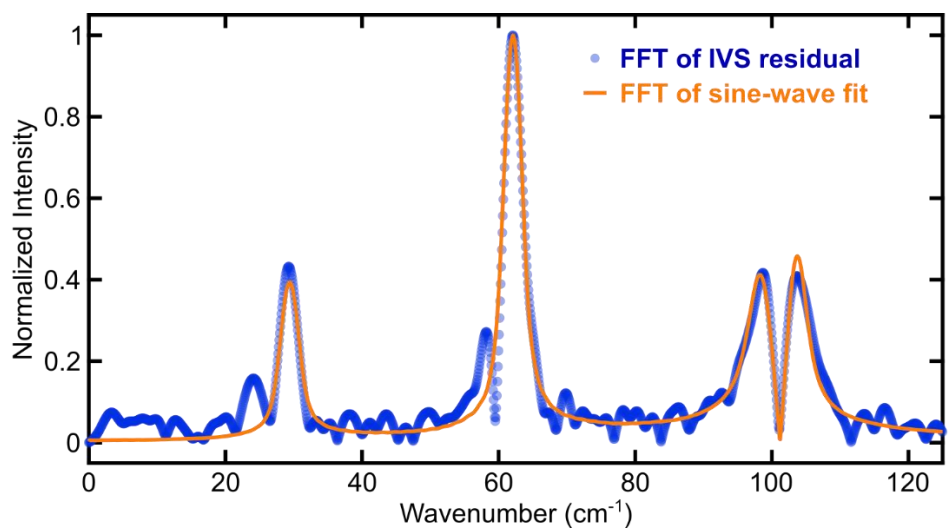

**Figure S7. Analysis of linewidth broadening mechanisms in IVS.** (blue) Fourier transform of IVS data at 5 K, identical to Figure 2d. (orange) Fourier transform of the sine-wave fit to the IVS data, shown in orange in Figure 2c.

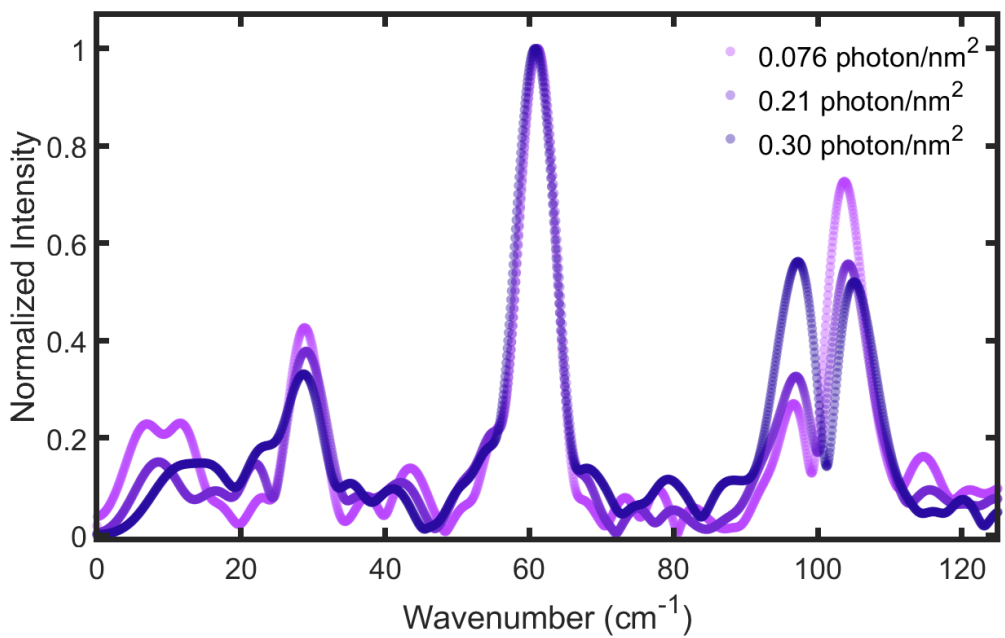

**Figure S8. Pump fluence-dependent IVS frequency domain spectra at 78 K.** The same analysis method was used on all three data sets, analyzing the 413-424 nm wavelength region on the blue edge of the highest energy ($X_3$) exciton. The highest fluence value of 0.30 photon/nm$^2$/pulse is the same fluence used to collect all other IVS experimental data presented in the main text, and this curve is the same as the 78 K curve in Figure 5d.

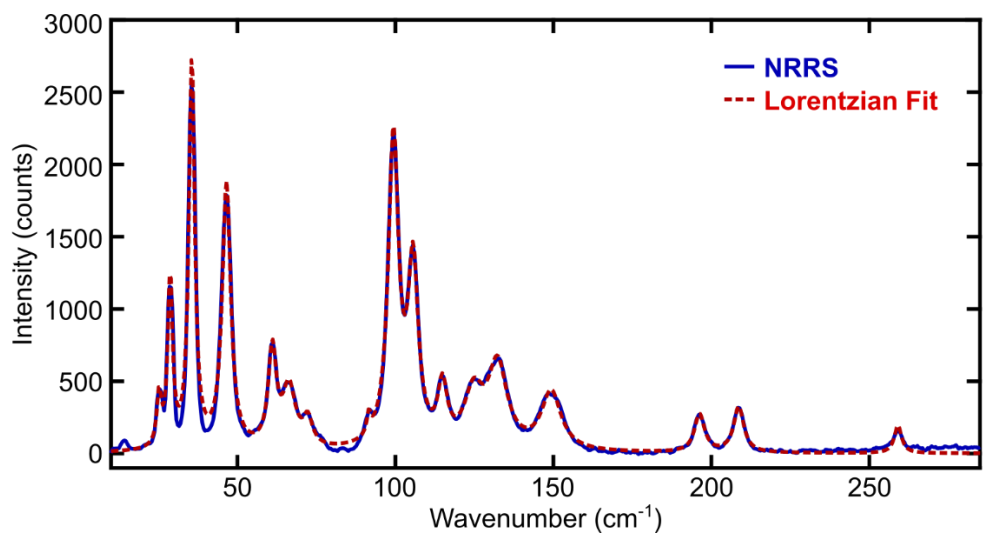

**Figure S9. Lorentzian fits to NRRS spectrum at 78 K.** Non-resonant Raman scattering spectrum (NRRS) at 78 K (blue) and the corresponding fit to multiple Lorentzian peaks (red).

## Section C: Vibrational Mode Data Analysis

**Supplementary Note 4: Converting Experimental and Computational Data into Correlation Scores**

The DFPT results contain a complete picture of all atomic displacements from equilibrium for each vibrational mode. This information was analyzed by evaluating the relative contributions from different sub-components to the total mode displacement. The categories considered included 1) the total contributions from the silver (Ag), selenium (Se), and phenyl ring (Ph) components, weighted by the amount of spatial displacement and the molecular weight of the component; 2) the changes in bond displacements for several key bonds in the structure, including $(Ag-Ag)_1$, $(Ag-Ag)_2$, Ag-Se, Se-C, and C-C; and 3) the degree of displacement for the Ag, Se, and Ph components projected along each of the three Cartesian coordinate directions. Each mode and category was evaluated independently from the others, with the contributions in each category summing to 1. The values for all of the various contributions to each mode are listed in Table S2 and Table S3.

The correlation scores reported in Figure 3d were found by calculating the degree to which each characteristic was correlated with IVS activity. For example, to calculate the score of 0.13 for the strongly correlated characteristic of Ag motion in the $z$-direction, we start with the Ag $z$-Motion Contribution column of Table S1. The values of this characteristic for the 5 IVS active modes ($\alpha$, $\beta$, $\gamma$, $\delta$, and $\zeta$) are summed:

$$0.35 + 0.06 + 0.07 + 0.09 + 0.57 = 1.14 \,. \tag{S1}$$

This value is divided by the total sum of all 9 modes in the column:

$$0.57 + 0.35 + 0.14 + 0.33 + 0.06 + 0.03 + 0.07 + 0.09 + 0.02 = 1.65, \tag{S2}$$

giving a value of

$$\frac{1.14}{1.65} = 0.69 \,. \tag{S3}$$

Finally, the value of a perfectly uncorrelated characteristic is subtracted:

$$0.69 - \frac{5}{9} = 0.69 - 0.56 = 0.13, \qquad (S4)$$

resulting in the reported Ag $z$-Motion correlation score of 0.13. This methodology gives a maximum correlation score of $1 - 0.56 = 0.44$ and a minimum correlation score of $0 - 0.56 = -0.56$.

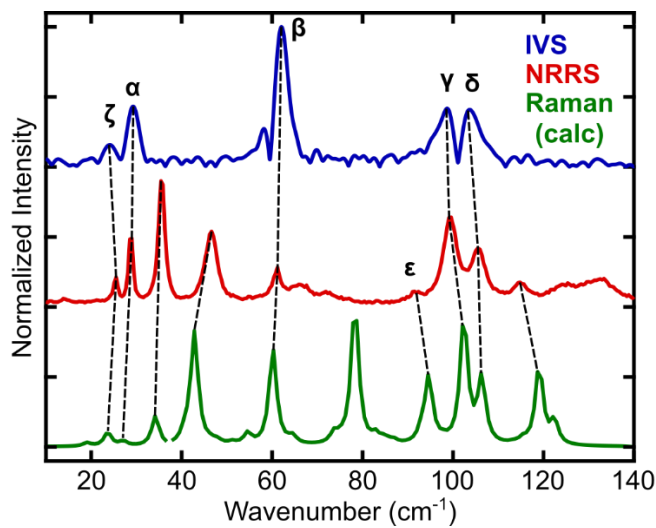

**Figure S10. Assignment of vibrational mode peaks between IVS, NRRS, and calculated Raman techniques.**

**Table S1. Summary of experimental and simulation results for all identified vibrational modes in AgSePh.**

| Mode Designation | NRRS at 78 K (cm$^{-1}$) | IVS at 5 K (cm$^{-1}$) | Calculated Raman [raw] (cm$^{-1}$) | Calculated Raman [shifted] (cm$^{-1}$) | IVS Lifetime at 5 K (ps) | Ag-Se Displacement Contribution | Ag $z$-Motion Contribution |
|---|---|---|---|---|---|---|---|
| Zeta (ζ) | 25.5 | 24.1 | 23.7 | 23.7 | - | 0.33 | 0.57 |
| Alpha (α) | 29.0 | 29.3 | 27.1 | 27.1 | 4.6 | 0.10 | 0.35 |
| - | 35.4 | - | 34.2 | 34.2 | - | 0.08 | 0.14 |
| - | 46.6 | - | 50.8 | 42.8 | - | 0.09 | 0.33 |
| Beta (β) | 61.2 | 62.1 | 68.2 | 60.2 | 4.7 | 0.41 | 0.06 |
| Epsilon (ε) | 91.5 | - | 102.7 | 94.7 | - | 0.02 | 0.03 |
| Gamma (γ) | 99.6 | 98.7 | 110.4 | 102.4 | 2.4 | 0.15 | 0.07 |
| Delta (δ) | 105.5 | 103.7 | 114.4 | 106.4 | 3.4 | 0.11 | 0.09 |
| - | | 114.8 | - | 127.1 | 119.1 | - | 0.37 | 0.02 |

**Table S2. Simulated atomic displacements and bond length changes for all identified vibrational modes in AgSePh at 5 K.** See also Figure S16.

| NRRS at 78 K (cm$^{-1}$) | Ag Contribution | Se Contribution | Ph Contribution | (Ag-Ag)$_1$ Displacement | (Ag-Ag)$_2$ Displacement | Ag-Se Displacement | Se-C Displacement | C-C Displacement |
|---|---|---|---|---|---|---|---|---|
| 25.5 | 0.85 | 0.10 | 0.05 | 0.45 | 0.20 | 0.33 | 0.01 | 0.01 |
| 29.0 | 0.62 | 0.15 | 0.23 | 0.30 | 0.55 | 0.10 | 0.02 | 0.03 |
| 35.4 | 0.40 | 0.18 | 0.42 | 0.28 | 0.52 | 0.08 | 0.03 | 0.09 |
| 46.6 | 0.61 | 0.19 | 0.20 | 0.28 | 0.61 | 0.09 | 0.01 | 0.02 |
| 61.2 | 0.29 | 0.25 | 0.46 | 0.03 | 0.48 | 0.41 | 0.02 | 0.07 |
| 91.5 | 0.07 | 0.33 | 0.60 | 0.39 | 0.06 | 0.02 | 0.18 | 0.35 |
| 99.6 | 0.39 | 0.36 | 0.25 | 0.35 | 0.42 | 0.15 | 0.01 | 0.07 |
| 105.5 | 0.22 | 0.22 | 0.57 | 0.19 | 0.06 | 0.11 | 0.04 | 0.61 |
| 114.8 | 0.42 | 0.33 | 0.26 | 0.31 | 0.16 | 0.37 | 0.02 | 0.14 |

**Table S3. Simulated atomic displacements along different Cartesian directions for all identified vibrational modes in AgSePh at 5 K.** See also Figure S16.

| NRRS at 78 K (cm$^{-1}$) | Ag $x$-Motion | Se $x$-Motion | Ph $x$-Motion | All $x$-Motion | Ag $y$-Motion | Se $y$-Motion | Ph $y$-Motion | All $y$-Motion | Ag $z$-Motion | Se $z$-Motion | Ph $z$-Motion | All $z$-Motion |
|---|---|---|---|---|---|---|---|---|---|---|---|---|
| 25.5 | 0.20 | 0.03 | 0.01 | 0.24 | 0.07 | 0.05 | 0.02 | 0.13 | 0.57 | 0.03 | 0.02 | 0.63 |
| 29.0 | 0.22 | 0.01 | 0.15 | 0.39 | 0.06 | 0.10 | 0.03 | 0.19 | 0.35 | 0.04 | 0.03 | 0.42 |
| 35.4 | 0.22 | 0.02 | 0.27 | 0.51 | 0.06 | 0.10 | 0.07 | 0.23 | 0.14 | 0.07 | 0.05 | 0.26 |
| 46.6 | 0.22 | 0.03 | 0.07 | 0.32 | 0.04 | 0.02 | 0.06 | 0.13 | 0.33 | 0.13 | 0.10 | 0.55 |
| 61.2 | 0.08 | 0.13 | 0.19 | 0.40 | 0.16 | 0.02 | 0.11 | 0.28 | 0.06 | 0.09 | 0.17 | 0.31 |
| 91.5 | 0.02 | 0.08 | 0.24 | 0.35 | 0.01 | 0.11 | 0.14 | 0.26 | 0.03 | 0.15 | 0.21 | 0.39 |
| 99.6 | 0.25 | 0.04 | 0.04 | 0.32 | 0.10 | 0.25 | 0.14 | 0.49 | 0.07 | 0.06 | 0.07 | 0.19 |
| 105.5 | 0.09 | 0.10 | 0.14 | 0.33 | 0.07 | 0.03 | 0.32 | 0.42 | 0.09 | 0.07 | 0.09 | 0.24 |
| 114.8 | 0.02 | 0.24 | 0.12 | 0.38 | 0.33 | 0.00 | 0.08 | 0.41 | 0.02 | 0.08 | 0.10 | 0.21 |

## Section D: Additional Temperature-Dependent IVS Figures

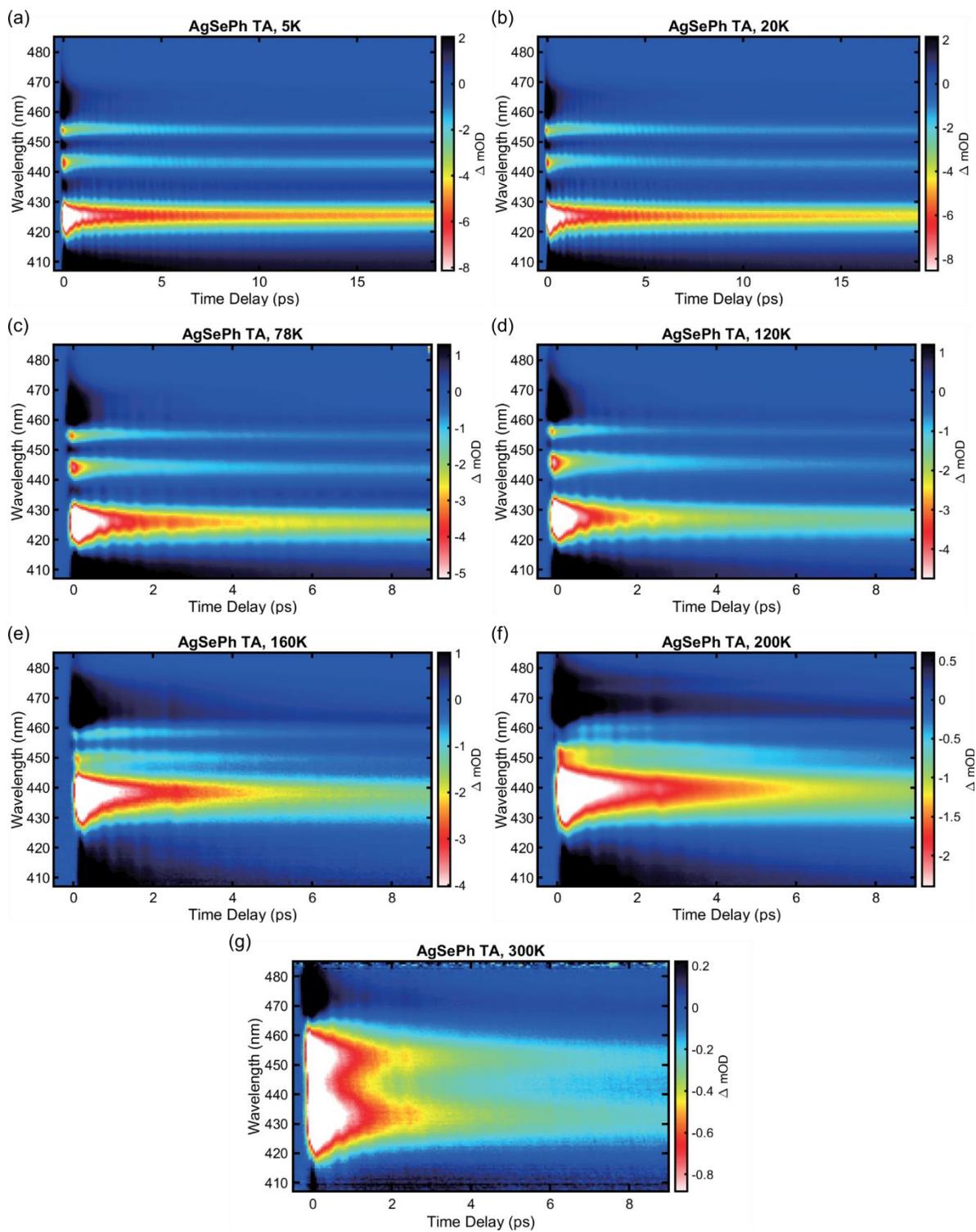

**Figure S11. IVS color maps of AgSePh measured at temperatures from 5 K to 300 K.**

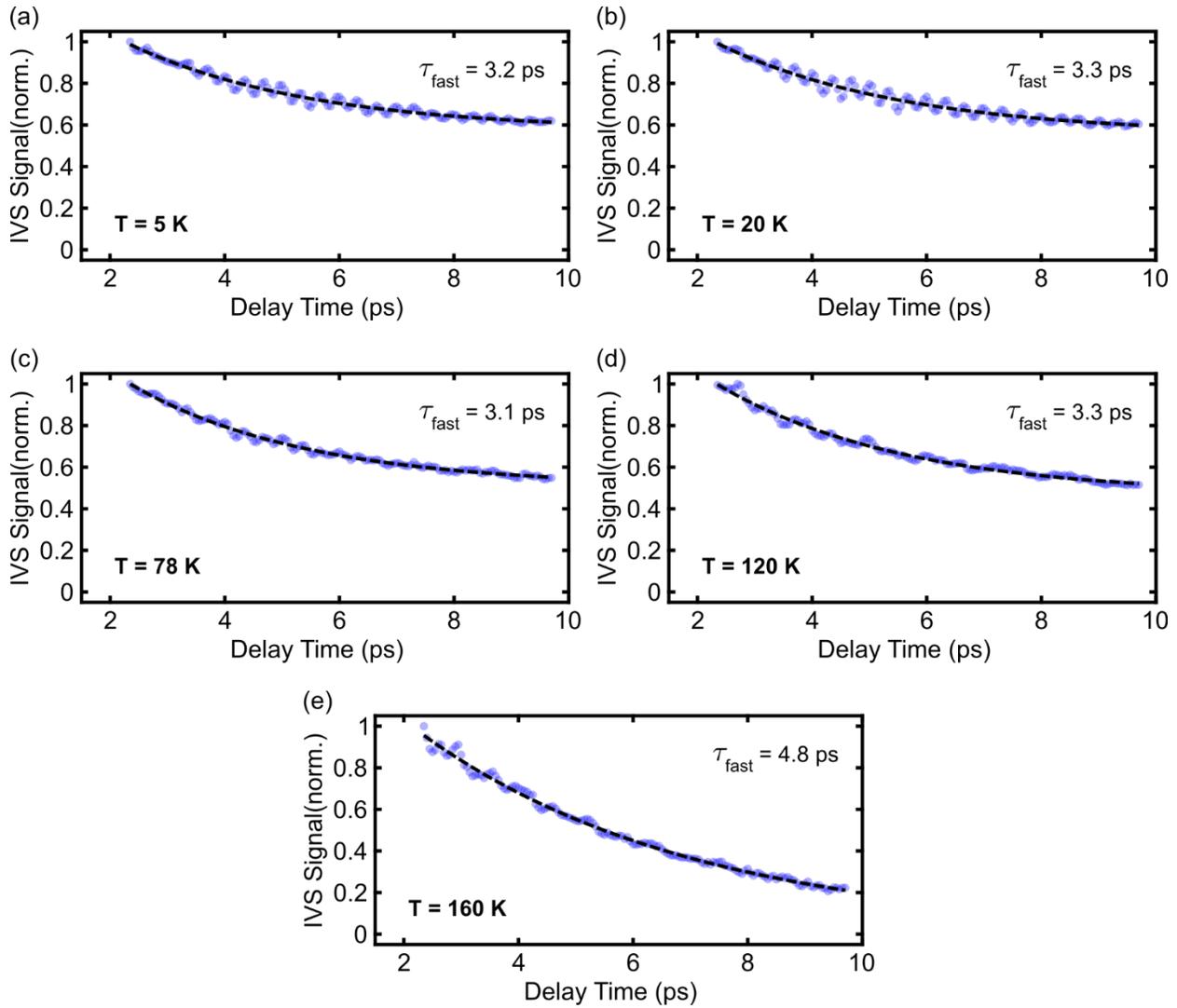

**Figure S12. Electronic decay dynamics of the $X_1$ bleach feature in AgSePh at temperatures from 5 K to 160 K.** The data was well fit with a biexponential decay, with the slow decay lifetime set to 1 µs. The fast lifetime is always found to be ~3-5 ps and doesn't show a clear trend with temperature, although the relative contribution of the fast lifetime component increases with increasing temperature. The small step in the data around 3 ps seen in some scans is due to a double reflection of the pump beam slightly re-exciting the sample. Data above 160 K was excluded due to increasing spectral overlap of the $X_1$ and $X_2$ bleach features.

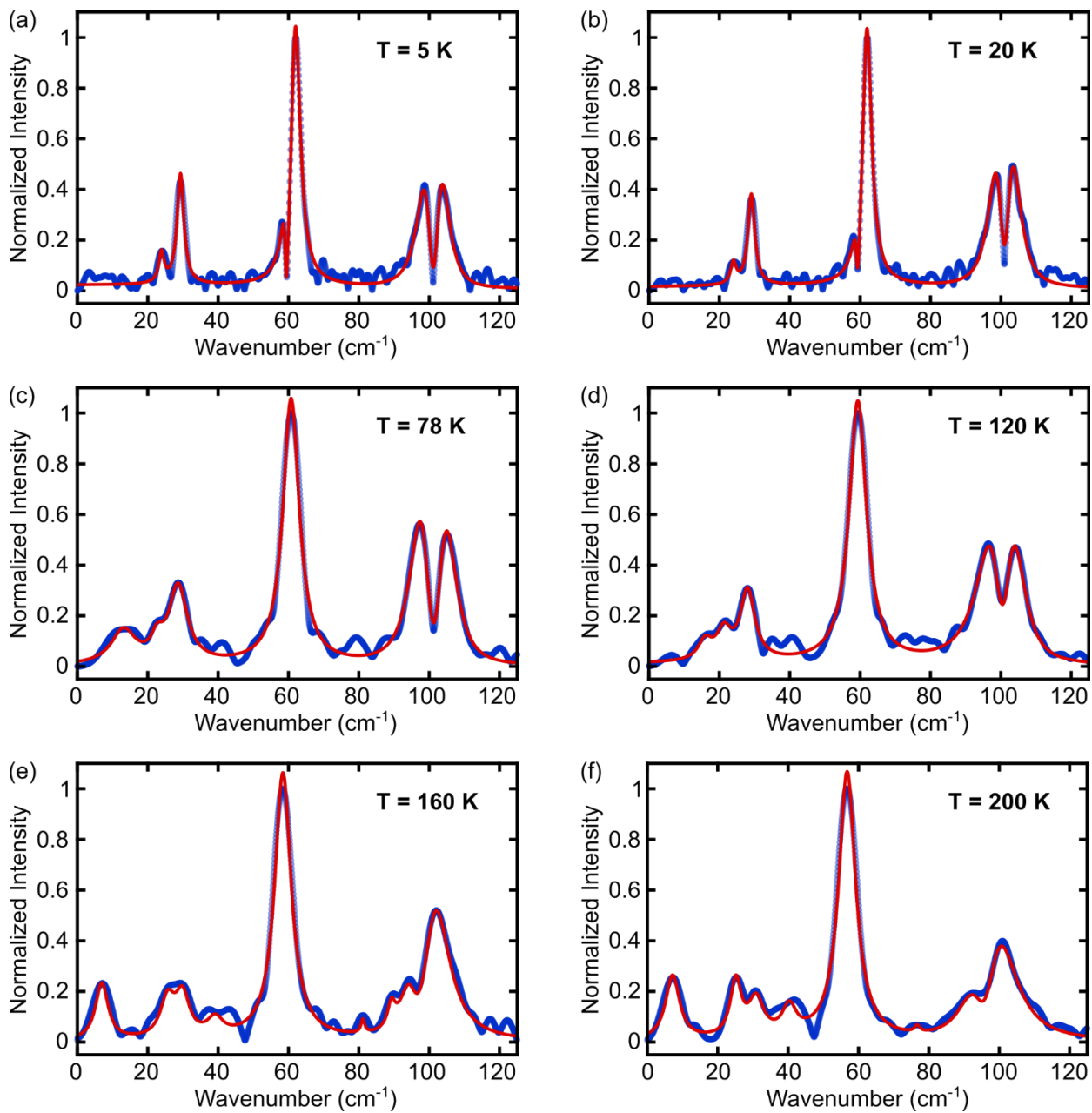

**Figure S13. Temperature dependence of the IVS spectrum from 5 K to 200 K.** These curves are equivalent to the data shown in Figure 5d, but separated and normalized at each temperature. The raw data is shown in blue, with a fit to the sum of multiple Lorentzian peaks in red.

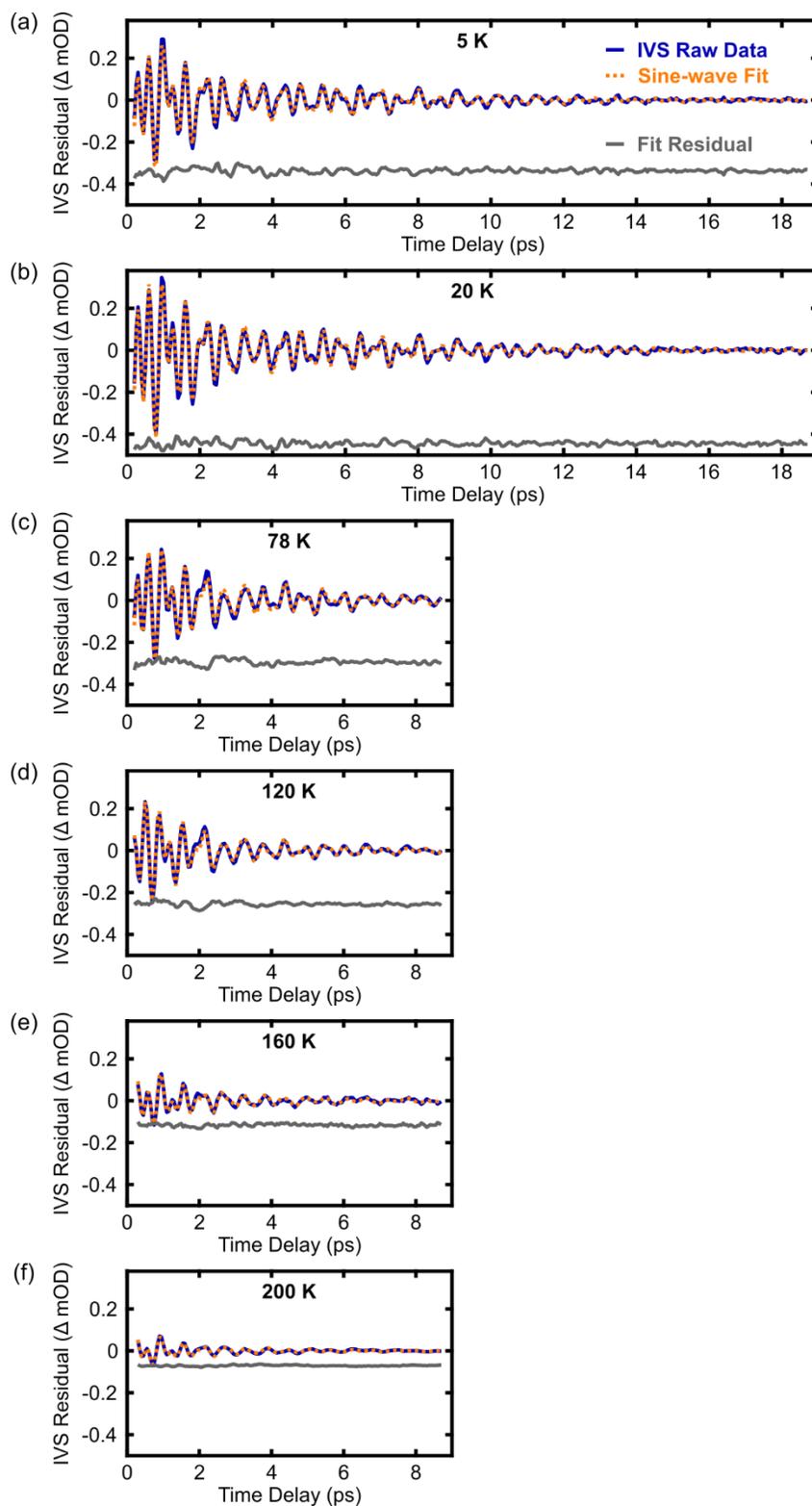

**Figure S14. Temperature-dependent IVS residuals with sine wave fits from 5 K to 200 K.** The raw data in blue is equivalent to that shown in Figure 5a. The orange curve represents the sine wave fit, with the extracted fit parameters shown in Table S4. The fit residual in gray is the difference between the experimental data and fit, offset for clarity.

**Table S4. Extracted parameters from time domain fits at all temperatures.**

| Designation | Parameter (units) | 5 K | 20 K | 78 K | 120 K | 160 K | 200 K |
|---|---|---|---|---|---|---|---|
| Alpha (α) | Frequency (cm$^{-1}$) | 29.4 | 29.4 | 29.0 | 27.8 | 27.0 | 26.4 |
| | Lifetime (ps)[a] | 4.6 | 5.2 | 4.9 | 4.2 | ∞ | ∞ |
| | Intensity (ΔmOD) | 0.7 | 0.6 | 0.5 | 0.4 | 0.1 | 0.0 |
| | Phase (deg) | 69 | 67 | 104 | 193 | 360 | 51 |
| Beta (β) | Frequency (cm$^{-1}$) | 62.1 | 62.0 | 60.6 | 59.4 | 58.7 | 56.9 |
| | Lifetime (ps) | 4.7 | 4.8 | 3.8 | 3.1 | 3.1 | 2.7 |
| | Intensity (ΔmOD) | 1.8 | 2.1 | 1.7 | 1.6 | 0.9 | 0.6 |
| | Phase (deg) | 213 | 208 | 236 | 298 | 327 | 6 |
| Gamma (γ) | Frequency (cm$^{-1}$) | 98.7 | 99.1 | 97.8 | 96.7 | - | - |
| | Lifetime (ps) | 2.5 | 2.8 | 1.9 | 1.3 | - | - |
| | Intensity (ΔmOD) | 2.2 | 2.6 | 2.5 | 3.1 | - | - |
| | Phase (deg) | 165 | 141 | 196 | 125 | - | - |
| Delta (δ) | Frequency (cm$^{-1}$) | 103.4 | 103.1 | 104.5 | 104.2 | 101.4 | 100.1 |
| | Lifetime (ps) | 3.4 | 3.0 | 2.8 | 2.2 | 1.2 | 1.0 |
| | Intensity (ΔmOD) | 1.3 | 2.6 | 1.2 | 1.0 | 2.6 | 1.5 |
| | Phase (deg) | 138 | 144 | 110 | 191 | 347 | 28 |

[a]*At 160 K and 200 K, the optimal fit to the time domain data exhibited no decay in the alpha mode over 10 ps, leading to the infinite decay lifetimes reported here.*

# Section E: DFT Computation Discussion and Figures

**Supplementary Note 5: Sources of Error in DFT Calculations**

A possible source of calculation error is the exchange correlation (XC) approximation. In this work, we used the PBE XC approximation, which provides a good trade-off between accuracy and computational cost. More advanced functionals, such as hybrids, could provide more accurate results, but at a prohibitively expensive computational cost for such a large electronic system. An additional complication is that the relatively complex crystal structure of the material, and the resultant large number of degrees of freedom, give rise to a "busy" theoretical spectrum with many modes of similar frequencies and intensities.

Some of the most noticeable differences between the experimental and calculated spectra can be found in the 150-350 cm$^{-1}$ frequency range. While the experimental non-resonant Raman spectrum shows some small peaks around 200 cm$^{-1}$ and 260 cm$^{-1}$, this is clearly quite different from the three high intensity split peaks that appear in the calculated spectra at approximately 180 cm$^{-1}$, 230 cm$^{-1}$, and 290 cm$^{-1}$ (Figure S15). When we examine the atom-projected spectra, we observe that only the Se and Ph components of the system contribute appreciably to these peaks. This indicates that vibrational motion of the Se and Ph components of the AgSePh system are suppressed in the experimental case relative to the theoretical case. Calculation error from the XC approximation's treatment of the $\pi - \pi$ between the phenyl groups could be one source of error[8], but there may also be a physical explanation for the suppression of these modes in the real system such as interactions between adjacent phenyl rings. Both the monolayer and bulk 2D AgSePh systems were found to have very similar spectra, and thus we conclude that inter-layer interactions are not responsible for suppressing these phenyl vibrations.

The shifted Raman spectrum (calc.) shown in Figures 3a and S10 matches all of the most prominent features in the experimental Raman spectrum, along with an additional Raman-active mode at 78 cm$^{-1}$ that does not appear in the NRRS or IVS data. We hypothesize that this calculated mode was not observed experimentally due to incompatibility between its symmetry and the geometry of our measurement (light propagating normal to the 2D plane).

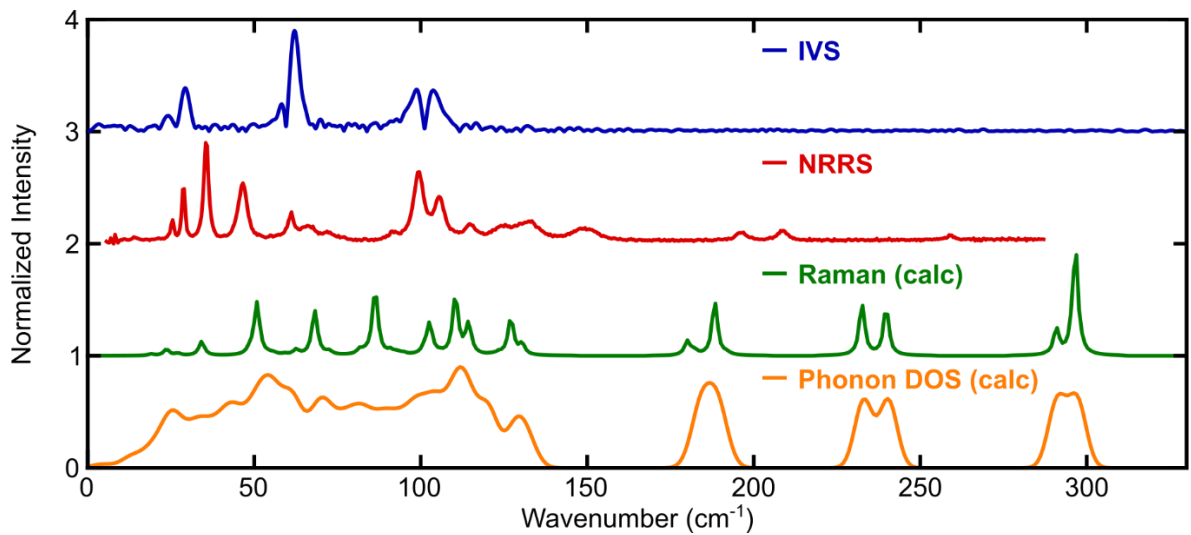

**Figure S15. Comparison between computational and experimental results over 0-350 cm$^{-1}$ frequency range.** The curves correspond to the same data shown in Figure 3a but are extended out to 350 cm$^{-1}$ to show higher frequency modes found in NRRS and computational results. The calculated curves are shown here without the 8 cm$^{-1}$ redshift correction applied in the main text.

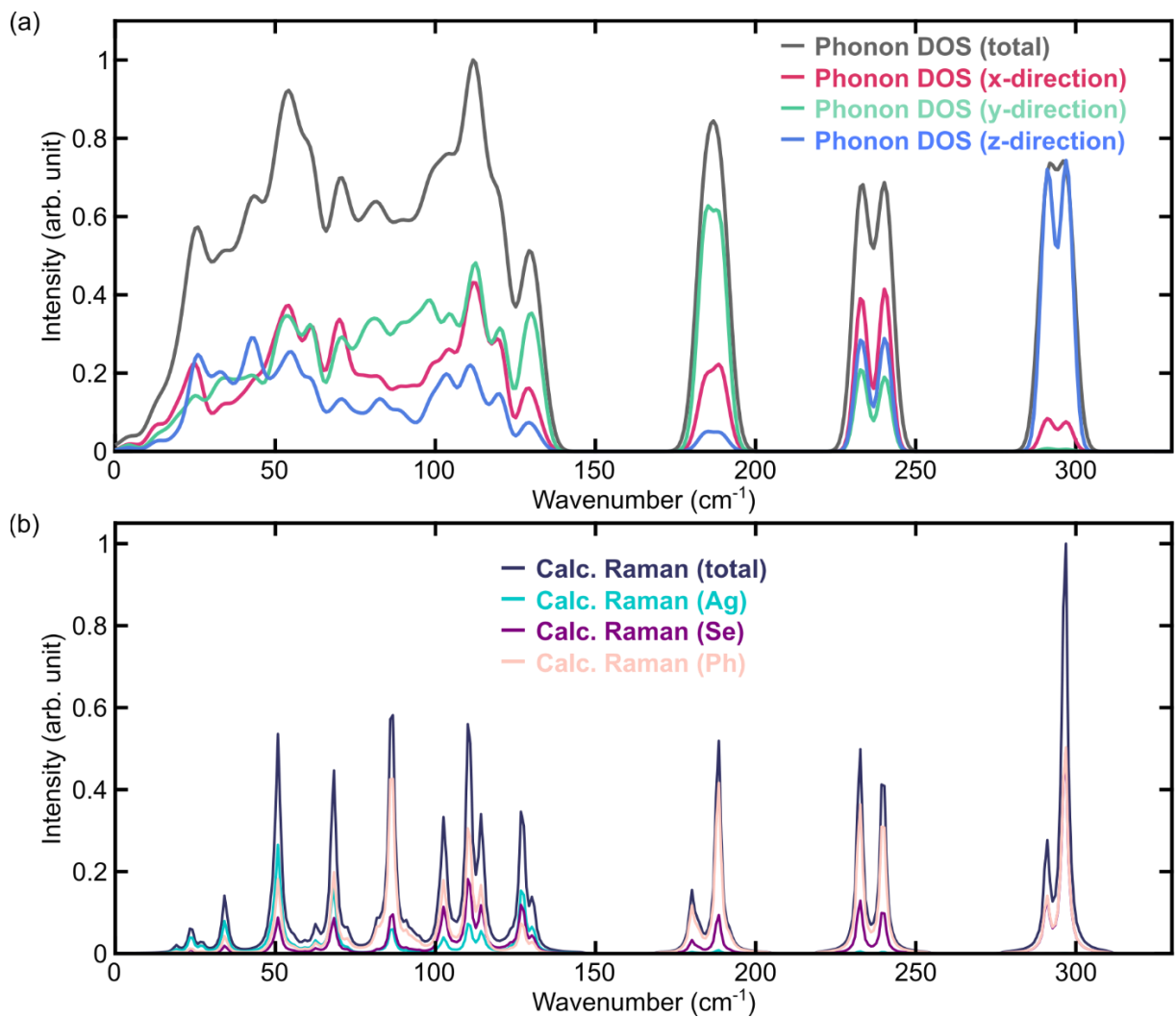

**Figure S16. Contributions from various vibrational mode characteristics to calculated Raman spectrum and phonon density of states.** The calculated curves are shown here without the 8 cm$^{-1}$ redshift correction applied in the main text.